\documentclass[aps,pre,showpacs,showkeys,superscriptaddress,%
preprintnumbers,a4paper,%
twocolumn,twoside,floatfix%
]{revtex4}

\usepackage{amsfonts,amsmath}
\usepackage{bm}
\usepackage[dvips]{graphicx}

\def\vec{\mathbf}

\def\cf{cf.~}
\def\ie{\emph{i.e.,} }
\def\eg{e.g.~}
\def\Re{\ensuremath{\mathcal{R}e}}

\def\rmd{{\mathrm{d}}}

\def\Fig#1{Fig.~\ref{fig:#1}}

\def\Tab#1{Table~\ref{tab:#1}}

\def\Eq#1{Eq.~(\ref{eq:#1})}

\def\sect#1{Sect.~\ref{sec:#1}}

\begin{document}

\title{Statistical analysis of coherent structures in transitional pipe flow}
\date{\today}

\author{Tobias M. Schneider}
\email{tobias.schneider@physik.uni-marburg.de} %
\affiliation{Fachbereich Physik, Philipps-Universit\"at Marburg, 
  Renthof 6, D-35032 Marburg, Germany}

\author{Bruno Eckhardt}
\email{bruno.eckhardt@physik.uni-marburg.de}
\affiliation{Fachbereich Physik, Philipps-Universit\"at Marburg, 
  Renthof 6, D-35032 Marburg, Germany}

\author{J\"urgen Vollmer}
\email{juergen.vollmer@physik.uni-marburg.de} %
\affiliation{Fachbereich Physik, Philipps-Universit\"at Marburg, 
  Renthof 6, D-35032 Marburg, Germany}

\begin{abstract}
  Numerical and experimental studies of transitional pipe flow have
  shown the prevalence of coherent flow structures that are dominated
  by downstream vortices. They attract special attention because they
  contribute predominantly to the increase of the Reynolds stresses in
  turbulent flow.  In the present study we introduce a convenient
  detector for these coherent states, calculate the fraction of time
  the structures appear in the flow, and present a Markov model for
  the transition between the structures.  The fraction of states that
  show vortical structures exceeds 24\% for a Reynolds number of
  about $\Re=2200$, and it decreases to about 20\% for $\Re=2500$.
  The Markov model for the transition between these states is in good
  agreement with the observed fraction of states, and in reasonable
  agreement with the prediction for their persistence. It provides
  insight into dominant qualitative changes of the flow when
  increasing the Reynolds number.
\end{abstract}



\maketitle 

\pagestyle{myheadings}
\markboth{T.M. Schneider, J. Vollmer, B. Eckhardt}%
{Coherent structures in transitional pipe flow}

\section{Introduction}

The visualization of turbulent flows and boundary layers via
sophisticated experimental methods like particle imaging velocimetry
has led to the identification of a rich variety of prominent coherent
structures, such as waves, streaks, hairpin vortices and lambda
vortices\cite{Robinson1991,Holmes1998,Podvin1998,Panton2001}. These
extended coherent structures significantly influence large scale
momentum transport and hence Reynolds stresses.  As a consequence,
they figure prominently in turbulence research.

Studies on internal flows in confined geometries have highlighted the
dominant role of structures containing pronounced downstream vortices
and have led to the proposal of a three-step self regenerating
mechanism for turbulence
\cite{Boberg1988,Jimenez1991,Trefethen1993,Hamilton1995,Waleffe1995,Waleffe1997,Waleffe1998,Grossmann2000,Jimenez2003}.
Downstream vortices transport liquid across the mean shear gradient
and create regions of fast or slow moving fluid, so-called high- and
low-speed streaks.  The streaks generated by this lift-up process
become unstable to the formation of normal vortices, and through a
nonlinear interaction mechanism the latter feed their energy back to
downstream vortices. This closed self-regeneration mechanism appears
to be a generic dynamical feature of turbulent shear flows.  The
process was identified in direct numerical simulations of plane
Couette flow in narrow cells where transverse modulations are
constrained \cite{Jimenez1991,Hamilton1995,Jimenez2003}, but it can
also be detected in time-correlation functions in fully turbulent
flows \cite{Jachens2006}.

In its purest form this self-regenerating cycle gives rise to a
periodic solution to the equations of motion. However, in most
coherent structures the flow is not strictly periodic and always
perturbed by background fluctuations. Examples of \emph{exact}
coherent states have been given in simple models, where they
correspond to periodic orbits
\cite{Moehlis2004,Moehlis2005,Smith2005,Smith2005b}, and, in the full
flow, through the numerical identification of exact coherent states in
channel flows
\cite{Nagata1990,Ehrenstein1991,Schmiegel1997,Nagata1997,Clever1997,Waleffe1998,Waleffe2001,Kawahara2001,Eckhardt2002}
and travelling waves in pipe flow \cite{Faisst2003,Waleffe2003,Wedin2004}.
In all these cases, the coherent structures are dominated by pairs of
counter-rotating downstream vortices and associated streaks which are
regularly arranged in azimuthal direction.  The flow fields are
invariant under discrete rotations around the pipe axis.

Since all exact coherent states constructed so far are linearly
unstable it came as a surprise that they could be directly observed in
experiments \cite{Hof2004}. In this work we follow up on this
experimental observation with a study of the appearance and
persistence of these structures in numerical simulations of pipe flow.
In particular, we show how they can be detected, how frequently they
appear, and how long they persist.

The travelling waves observed in pipe flow are of particular interest
because they are believed to form a backbone for the turbulent
dynamics near the onset of turbulence.  Since the laminar profile in
these flows is linearly stable for all Reynolds numbers
\cite{Davey1969,Salwen1980,Patera1981,Drazin1981,Brosa1986,Brosa1989,Herron1991,Meseguer2003}
the transition cannot proceed through states bifurcating from the
laminar profile. The turbulent motion which in many pipe-flow
experiments is observed for Reynolds numbers beyond about 2000, must
hence arise via a nonlinear transition scenario
\cite{Boberg1988,Waleffe1995,Brosa1999,Eckhardt1999,Faisst2004,Eckhardt2007}.
The travelling waves are then the simplest persistent nonlinear structures
around which the turbulent dynamics can form. Together with their stable
and unstable manifolds they can give rise to the basic building blocks
of chaotic dynamics, such as hyperbolic tangles and Smale
horseshoes. While it is unlikely that one will be able to identify an 
individual travelling wave in a time series, it is possible to identify a
visit to their neighborhood, as identified by the appearance of similar
structures in the flow. 

In the present paper we propose a way to detect the visits to the
neighborhoods of coherent states, and use it to infer information
about the structures underlying turbulence.
To distinguish different parts of state space and different flow
topologies, we introduce projections onto lower dimensional subspaces
that capture salient features of classes of coherent states, and study
the recurrences to these subspaces: This is weaker than identifying
individual travelling waves but sufficient to discriminate between
various flow regimes.  On the technical side, the reduction in
resolution also lowers the requirements on the length of the time
traces and helps to improve the statistical significance.

The outline of the paper is as follows. 
In \sect{numerics} we briefly describe the spectral code underlying
the simulation of the flow.  
In \sect{detection} we describe the projection used to detect and
characterize the coherent structures in direct numerical simulations
of pipe flow close to the threshold of turbulence.  
In \sect{time-series} we analyst the statistics of the occurrence of
coherent structures, and in \sect{physProp} we explore their physical
properties.  The paper closes with a discussion and outlook in
\sect{discussion}.

\section{Simulation of pipe flow}
\label{sec:numerics}

We consider an incompressible Newtonian liquid in a pipe of circular
cross section subject to no-slip boundary conditions at the walls.
The flow is forced by a uniform pressure gradient which is adjusted to
keep the flux constant at any instant of time
\cite{Draad1998,Darbyshire1995,Hof2004}.  In other words the
integrated volume flux through a cross section of the pipe is
constant, and the Reynolds number
\begin{equation}
  \Re = \frac{2 \langle u_z\rangle R}{\nu}
\end{equation}
is externally controlled in order to be independent of the flow state
of the liquid.
Here $\langle u_z\rangle$ denotes the mean downstream velocity, $R$ is
the pipe radius and $\nu$ the kinematic viscosity of the liquid.  In
our simulations the pipe is $L=10 R$ long, and we use periodic
boundary conditions in the downstream direction: physically, this
corresponds to a numerical representation of the interior of a
turbulent patch.

The Navier-Stokes equations are written in cylindrical coordinates
$(r,\varphi,z)$ and solved with a pseudo-spectral scheme.  In doing so
dimensionless units where lengths are measured in units of the radius
of the pipe and velocities in units of twice the mean downstream
velocity (\ie the centerline velocity of the equivalent parabolic
laminar profile) are used. Time is measured in units of 
$R / 2 \langle u_z \rangle$.

All three components of the velocity field $(u_r, u_\varphi, u_z)$ are
decomposed into Fourier modes in azimuthal and downstream direction.
Chebyshev polynomials are used for expansion in radial direction.  The
velocity field is thus written as
\begin{equation}
 \left(
\begin{aligned}
u_r\\
u_\varphi\\
u_z
\end{aligned}
\right) = \sum_{n, m, j}
 \phi_{n, m, j}
\left(
\begin{aligned}
c_r\\
c_\varphi\\
c_z
\end{aligned}
\right)_{n,m,j} \, .
\end{equation}
Here the spectral basis functions are 
\begin{equation}
\phi_{n,m,j} (r,\varphi, z) 
\equiv 
\frac{1}{2 \pi L} \: \mathrm{e}^{i(n \varphi + m k_z z)} \: T_j(r),
\end{equation}
where $T_j$ denotes the $j^{th}$ normalized Chebyshev polynomial
\cite{Abramowitz1984,Canuto1988}, and $k_z = \frac{2 \pi}{L}$.  In
physical collocation space the velocity fields are represented by the
values of the fields at the corresponding Gauss-Lobatto grid points.

A 4th-5th-order Runge-Kutta-Fehlberg scheme with adaptive step-size
control is used to evolve the solution in time
\cite{NumericalRecipes77}, and the action of the
Navier-Stokes operator is computed via a pseudo-spectral scheme.  The
transformation between spectral and physical space required by the
pseudo-spectral scheme is performed by FFT-based routines. Constraints
(incompressibility, regularity and analyticity) as well as no-slip
boundary conditions are enforced by a Lagrangian projection mechanism
\cite{Schneider2005}.

The simulations presented in this work are carried out with $n$
Fourier modes in azimuthal and $m$ Fourier modes in downstream
direction, where $\frac{|n|}{24} + \frac{|m|}{22} \le 1 \,$.
Consequently, we consider up to $49$ Fourier modes in azimuthal and up
to $45$ in the downstream direction. $47$ Chebyshev polynomials are
used for the expansion in the radial direction, adding up to 
$3\times 49\times 23\times 47\approx1.6\times 10^5$ components.

%
\begin{figure}
\[ \includegraphics[width=0.45\textwidth]{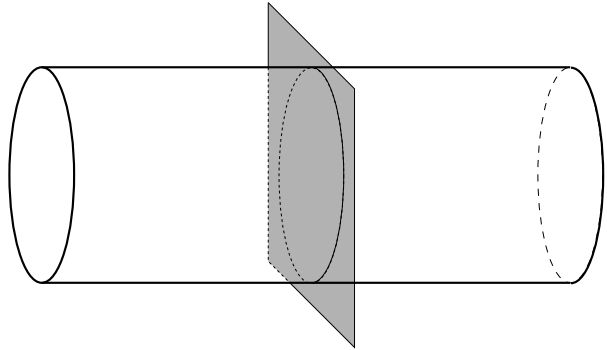} \] 
\caption[]{
  The coherent vortices are expected to be 
  most prominent in cross sections perpendicular to the 
  pipe axis, as the shaded plane  
  in this figure. In experiments, the velocity fields in
  this cross section 
  are obtained by steroscopic particle image
  velocimetry 
  \cite{Hof2004}. 
  \label{fig:geometry}}
\end{figure}
%

\begin{figure}
\[
\parbox[b]{2mm}{\textrm{(a)}\\[0.17\textwidth]} 
\includegraphics[height=0.18\textwidth,clip]{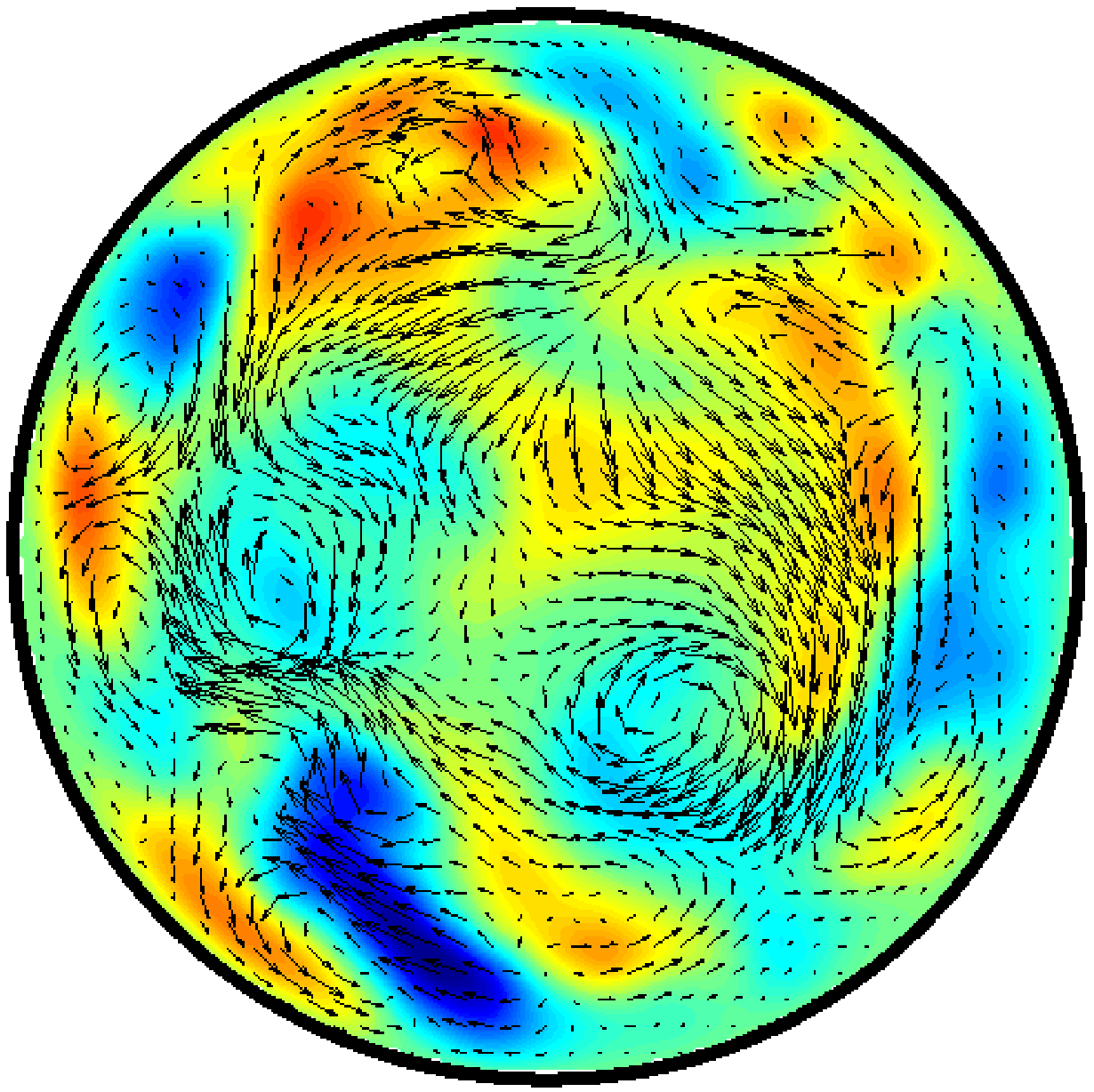}
\parbox[b]{5mm}{\ \\[0.21\textwidth]} 
\parbox[b]{2mm}{\textrm{(b)}\\[0.17\textwidth]} 
\includegraphics[height=0.18\textwidth,clip]{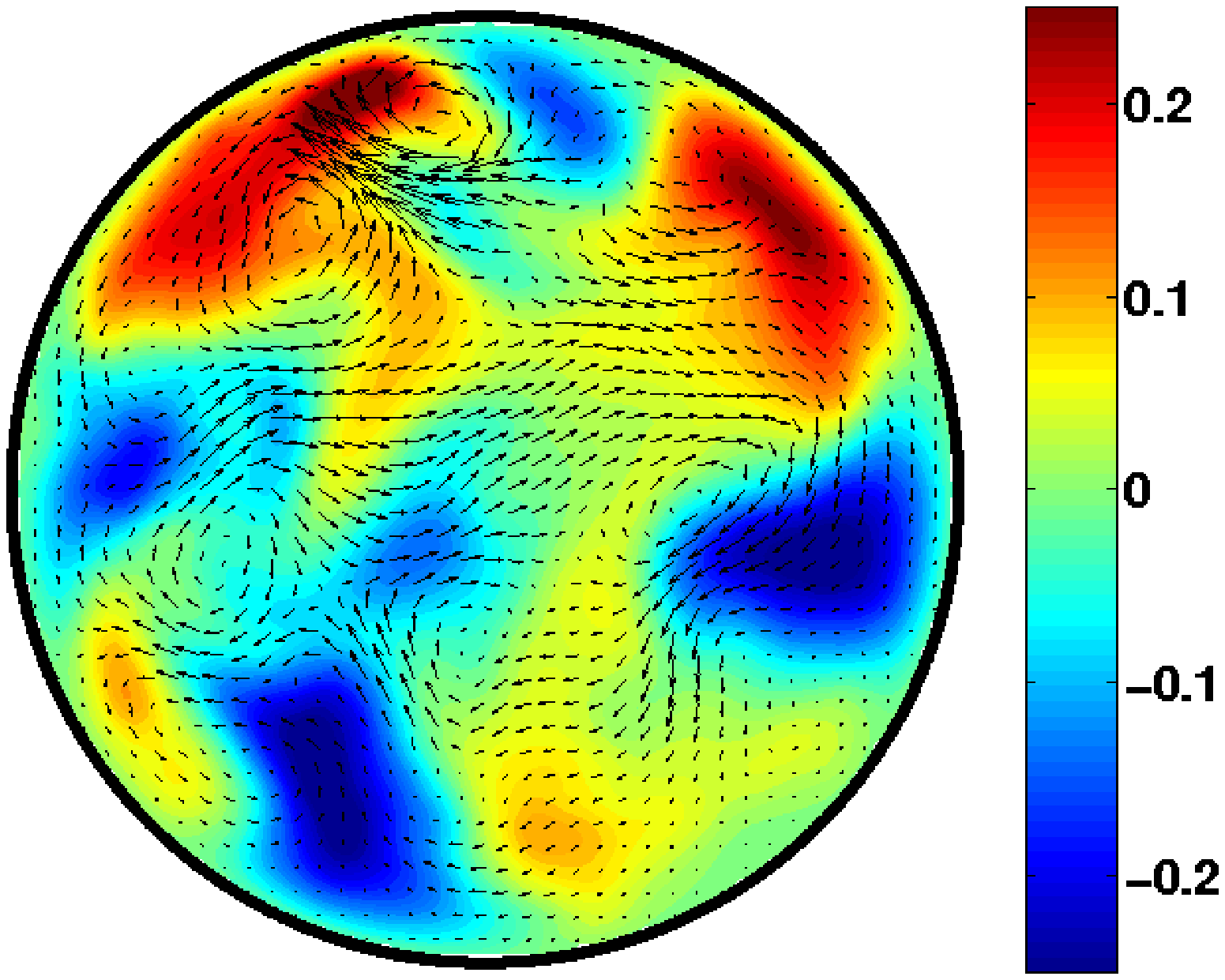}
\]
\caption[]{(color online) 
  Deviation $\vec{u} - \langle \vec{u}\rangle_{t}$ of the
  instantaneous velocity field $\vec{u}$ from the mean turbulent
  profile $\langle \vec{u}\rangle_{t}$ for a pipe flow at 
  $\mathrm{\Re}=2200$. The colors indicate the downstream velocity 
  component according to the scale specified by the color bar 
    to the right, and the in-plane velocity components are 
  indicated by arrows.
  The two panels 
  show (a) a case where no clear structure is observed,
  and (b) one with a four-fold streak. 
  \label{fig:crossections}}
\end{figure}

\section{Detection of Coherent Structures}
\label{sec:detection}

The travelling waves \cite{Faisst2003,Wedin2004} we want to detect are
dominated by vortices aligned along the axis, and corresponding
streaks in the downstream velocity components.  The downstream
vortices and streaks are most prominent in cross sections of the pipe
perpendicular to the axis. As in the experiments \cite{Hof2004}, where
stereoscopic particle image velocimetry was used to extract the
velocity fields, we will focus on the velocity fields in cross
sections perpendicular to the pipe axis (\Fig{geometry}).  For the
travelling waves it makes no difference whether we focus on one cross
section and follow the time evolution or whether we freeze the flow at
one instance of time and move the cross section along the axis. The
same applies for a transient appearance of these structures: in a
fixed cross section they will come and go, and in a frozen flow they
would be present in some regions along the axis and absent in others.
In the analysis presented below we work, as in the experiments, with
the time evolution in cross sections at a fixed position in the lab
frame. Typical examples of cross sections with high- and low-speed
streaks, \ie of regions of high and low downstream velocity, are shown
in \Fig{crossections}. The structures are best visible when a
reference profile is subtracted. In previous works \cite{Hof2004} the
laminar profile with equal mean velocity was subtracted.  Here we use
the mean turbulent profile.  It is obtained as the average over
azimuthal angle and time of the downstream velocity at a fixed radius.

\subsection{Characterizing the symmetry of coherent states}

\begin{figure}
\[
\includegraphics[width=0.20\textwidth]{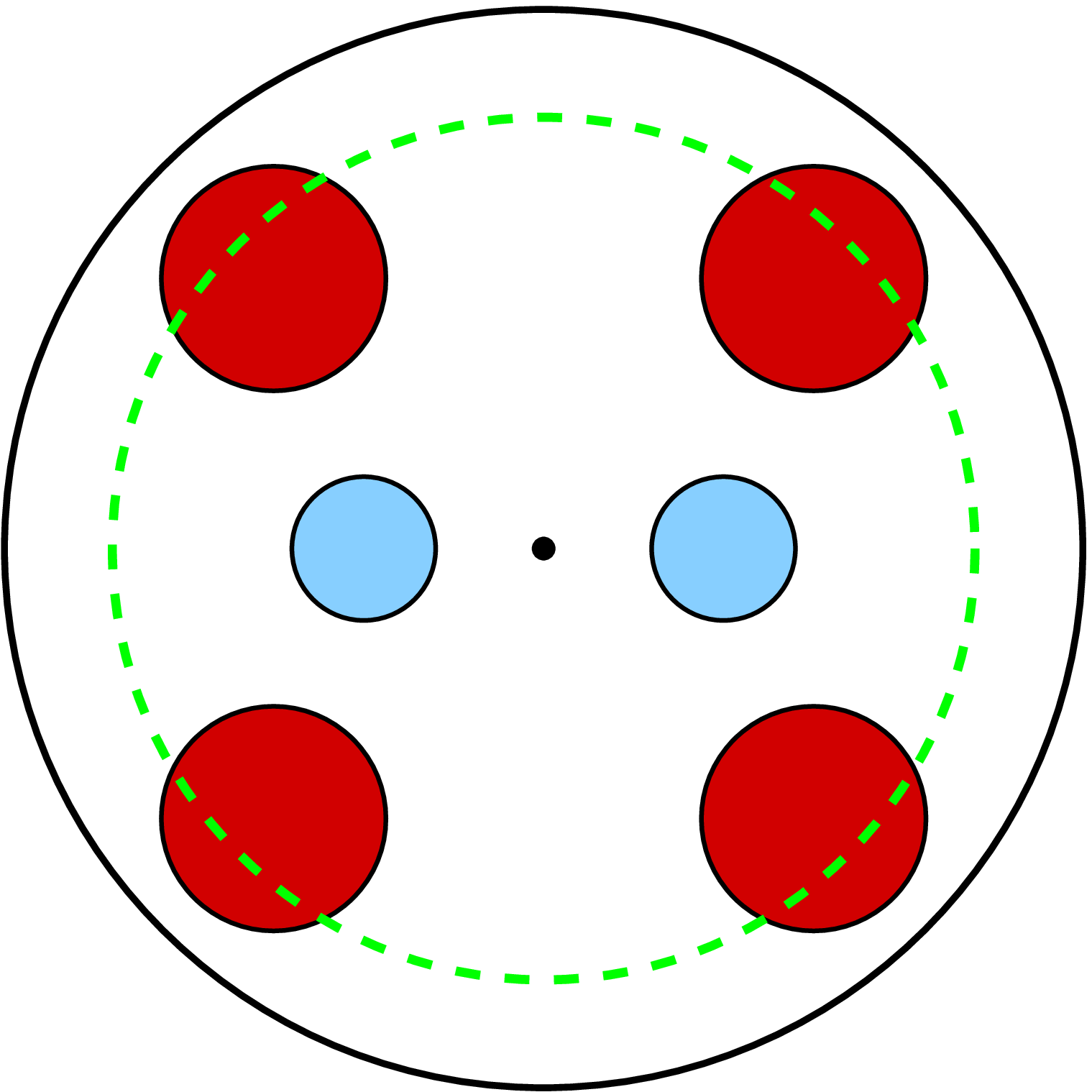}
\quad\quad
\includegraphics[width=0.20\textwidth]{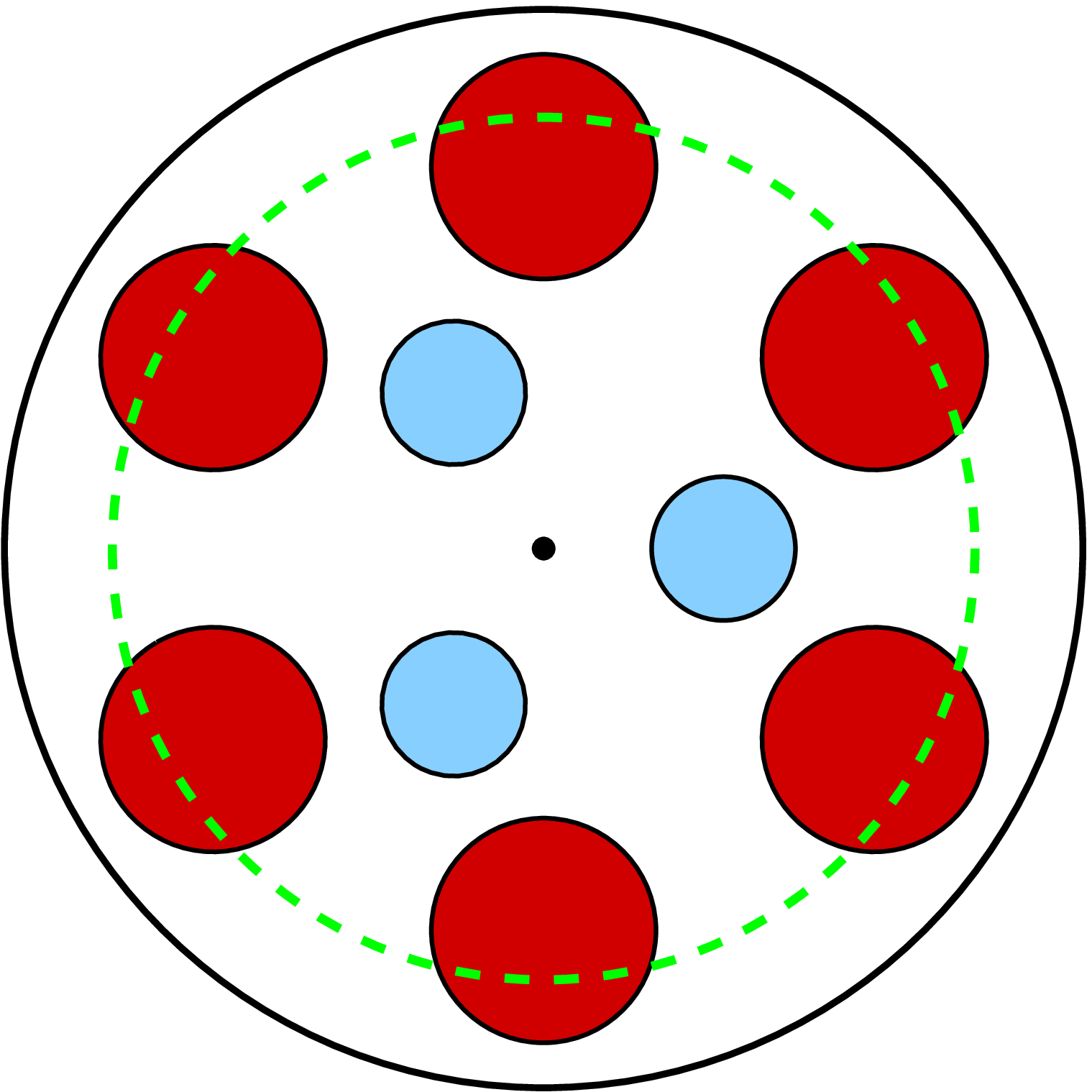}
\]
\caption[]{(color online) 
  Sketch of the regular arrangement of high- (dark red) and
  low-speed (light blue) streaks in coherent structures. 
  When analyzed at a fixed radial position close to the wall 
  (green dashed line at radius $0.81$), all currently known
  travelling-wave solutions show high-speed streaks that are
  equidistantly arranged on the circumference, \ie they show an
  $N$-fold rotational symmetry. Typical states contain $N$ low speed
  streaks close to the center, and $2N$ high speed streaks close to
  the wall \cite{Faisst2003}. However, also states containing $N$ low
  and $N$ high speed streaks were found \cite{Wedin2004}.
  \label{fig:correlators}}
\end{figure}

As mentioned in the introduction the coherent travelling waves
identified so far have highly symmetric arrangements of vortex pairs.
By transporting fast liquid from the center to the walls and slow
liquid from the wall to the center region, these pairs of vortices
generate elongated regions of fast and slow moving liquid. We
therefore focus on the appearance of symmetric arrangements of high-
and low-speed streaks schematically indicated in \Fig{correlators}.
The travelling waves also show that the high-speed streaks close to
the walls are fairly stable and do not move much in the azimuthal
direction over one period. This simplifies their detection amidst the
fluctuations of the total velocity field.

The rotational symmetry of the pipe entails that patterns should be
considered identical when they only differ by a global rotation around
the pipe axis. A detector for coherent states should take this into
account and be invariant under global rotations.  We therefore suggest
to use an azimuthal correlation of the downstream velocity $u_z$ at a
chosen radius $r$ and axial position $z$,
\begin{equation}
  C(\phi) 
\equiv 
\langle u_z\left(r, \varphi, z\right) \; 
        u_z\left(r, \varphi + \phi, z\right) 
\rangle_\varphi\;,
\label{eq:correlationFunc}
\end{equation}
where $\langle\cdot\rangle_\varphi$ denotes averaging over $\varphi$.
By a straightforward calculation one verifies that this correlation
function is invariant under global rotations. Moreover, it reliably
uncovers periodic structures in the azimuthal direction.

\begin{figure}
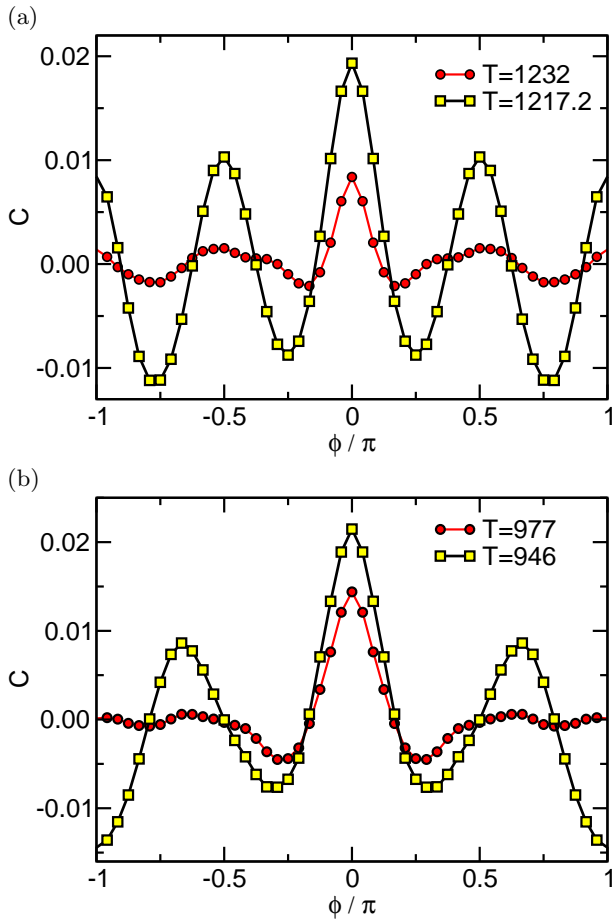

\flushleft{(a)}\\
\includegraphics[width=0.45\textwidth]{schneider_Fig4a}
\\
(b)
\\
\includegraphics[width=0.45\textwidth]{schneider_Fig4b}
\caption[]{(color online) 
  Azimuthal correlation functions evaluated at $r=0.81$ for
  the velocity fields shown in \Fig{crossections}.  When no clear
  structure is observed in the cross section, the correlation function
  only shows an autocorrelation peak at $\phi = 0$ (red circles). 
(a) When a four-fold symmetry is present, \eg for $t=1217.2$,
  the correlation functions 
  has additional peaks at $\phi = \pm \frac \pi 2$ and $\pi$ (yellow boxes). 
 (b) For a three-fold symmetric state, the additional peaks appear 
 at $\phi = \pm \frac{2\pi}{3}$.
 \label{fig:correlations}}
\end{figure}

Whenever the system approaches a coherent state showing $N$
high-speed streaks close to the wall, the 
correlation function 
$C(\phi)$ shows $N$ peaks separated by an angular displacement 
${2 \pi}/{N}$. 
In particular, 
the four-fold structure of the downstream velocity field, 
\Fig{crossections}b results in a clear four-fold structure of the
correlation function, 
which is shown in \Fig{correlations}a. 
In addition to the autocorrelation peak at $\phi=0$ 
the correlation function shows peaks at $\phi = \pm \frac \pi 2, \pi$. 
Similarly, a flow with a three-fold symmetry gives rise to 
peaks at $\phi = 0, \pm \frac{2\pi}{3}$
(\cf\Fig{correlations}b).

\begin{figure}
\[
\includegraphics[width=0.45\textwidth]{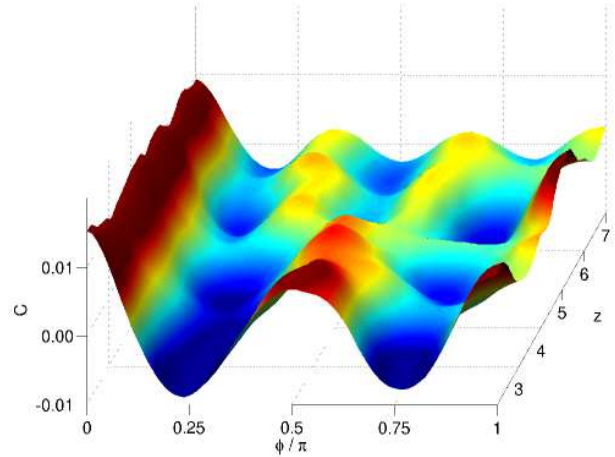}
\]
\caption[]{(color online) 
  Azimuthal correlation function plotted as a function of
  downstream position in the pipe. One clearly observes the
  transition form a four-streak state to a six-streak state. The
  transition is quite sharp and happens within a spatial range of a
  single pipe radius.
  \label{fig:transition}}
\end{figure}

By following $C(\phi)$ in time one can detect the lifetimes of
structures, their decay, and the subsequent emergence of new patterns.
An example is given in \Fig{transition}, which shows the decay of a
four-streak state and the emergence of a six-streak state within about
$1$ pipe radius.

\begin{figure*}
  \begin{minipage}{0.49\textwidth}
    \begin{center}
      \includegraphics[width=0.98\textwidth]{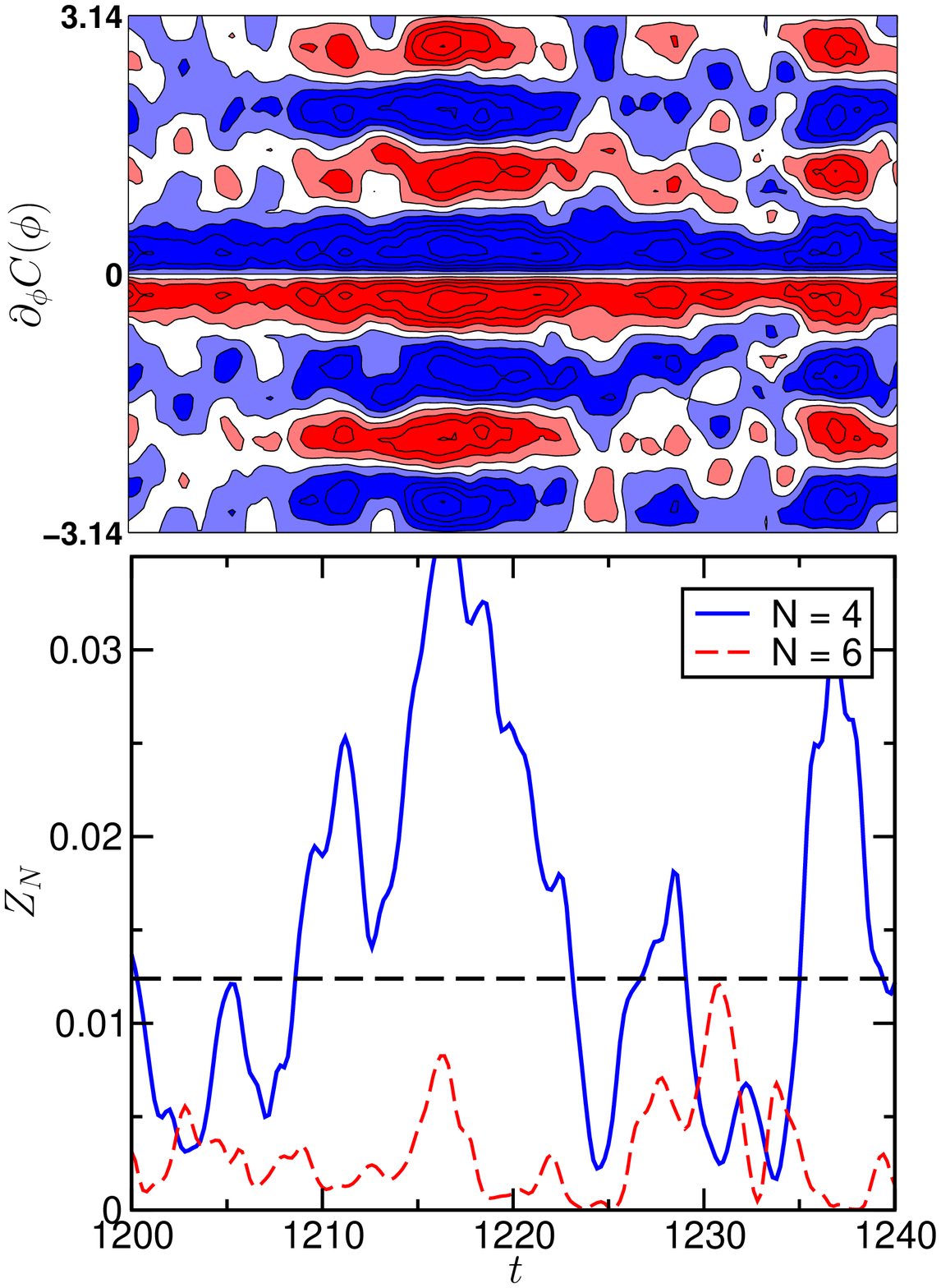}
    \end{center}
  \end{minipage}
  \begin{minipage}{0.49\textwidth}
    \begin{center}
      \includegraphics[width=0.98\textwidth]{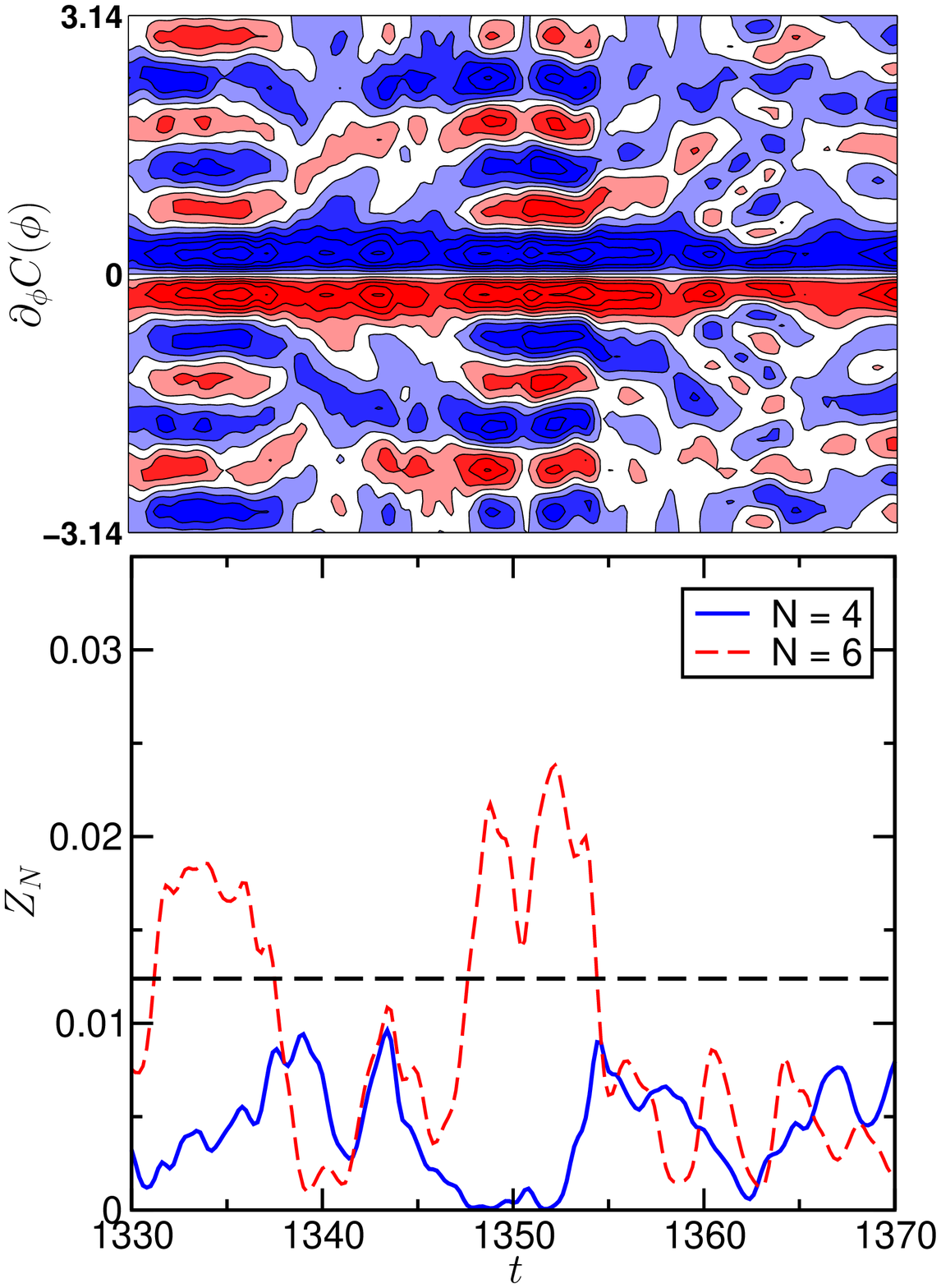}
    \end{center}
  \end{minipage}
  \caption[]{(color online) 
    The derivative of the correlation function 
    $\partial_\phi C(\phi)$ 
    as a function of $\phi$ and time $t$ 
    (top), and of the corresponding scalar
    measures $Z_4$ (blue solid line) and $Z_6$ (red dashed line) in
    bottom part. The color coding in the top graphs runs linearly from
    $-0.005$ (blue) to $0.005$ (red). Nodal lines appear in white.
    Besides irregular, featureless correlation functions at \eg
    $t=1200\ldots1208$ and around $t=1230$, there are long stretches
    of time where the function shows a distinct four-fold (\eg at
    $t=1210\ldots1220$) and six-fold symmetry (\eg around $t=1334$ and
    $1350$), respectively.
    \label{fig:scalarmeasure}}
\end{figure*}

\subsection{Automated structure detection}

The correlation function $C (\phi)$ signals the proximity of the
  flow to a coherent state by evenly spaced peaks.  Its derivatives
  highlight both minima and maxima of the correlation function (see
  \Fig{scalarmeasure}) and emphasize flow structures of comparable
  (azimuthal) streak gradients. Since $C (\phi)$ is an even function
  in $\phi$, its derivative is odd. It
  should have an substantial overlap with the sine-function of the
  appropriate periodicity.
In order to automatically detect evenly spaced maxima and in order to
count their number we therefore define the scalar measures $Z_N$ via a
scalar product of the derivative of the correlation function and
$\sin(N \phi)$ \footnote{The scalar measure can be effectively
  computed as
$ Z_N = N \int_{-\pi}^{\pi} C(\phi) \cos(N \phi)\; d \phi\; $.
}, 
\begin{equation}
  Z_N(t)
  \equiv 
  - \int_{-\pi}^{\pi} \partial_\phi C(\phi) \sin(N \phi)\; \mathrm{d} \phi\;.
\end{equation}
This reduction of information to a scalar quantity contains one
parameter, the radius $r$ at which the correlation functions are
determined. For the Reynolds numbers considered here we find $r=0.81$
to be convenient. At this radius, which is indicated by a dashed green
line in \Fig{correlators}, the coherent structures under investigation
show a pronounced regular arrangement of high-speed streaks.

By following time traces of $Z_N$ for different $N$ we can study the
prevalence of structures of certain multiplicity and the transitions
between them. Examples are given in \Fig{scalarmeasure}.  The top
frames show $\partial_\phi C(\phi)$, the derivative of the azimuthal
correlator with respect to the angular coordinate $\phi$, as a
function of the azimuthal coordinate $\phi$ and the time $t$.  The
four-fold structures have eight zeros in their derivative (from four
maxima and four minima), and the six-fold structures have twelve
zeros. Parallel nodal lines indicate the presence of these structures
for times of about ten natural time units.

The lower frames in \Fig{scalarmeasure} show the time evolution of the
corresponding scalar projectors $Z_N$.  The indicator $Z_4$ shows 
pronounced peaks when the four-fold symmetric patterns are observed 
in the correlation function and $Z_6$ peaks when the six-fold structures
appear; conversely, one is small when the other one is large. One also
notes considerable fluctuations due to the residual background
turbulence. In general, values of $Z_N$ smaller than about $0.01$
cannot be considered significant indicators of a structure and belong
to background fluctuations. On the other hand, comparison of the top
and bottom frames in \Fig{scalarmeasure} suggests that a threshold
$
  Z_N > 0.013
$
signifies the presence of coherent structures with
$N$-fold symmetry.

Armed with this threshold, we collapse the scalar time series $Z_N(t)$
for $N=2,\ldots,8$ to a single discrete indicator, $N(t)$, which
assigns to a cross section at time $t$ the number $N$ of symmetric
streaks and corresponding vortices it contains. $N$ takes the values
$0$, $2$, $3, \ldots ,8$, where $N=0$ is assigned to cases where all
$Z_N$ remain below the threshold.  The maximal value $8$ is an
empirical limit, in that states with eight or more vortices were
rarely realized.

\begin{figure*}
  \[
  \parbox[b]{2mm}{(a)\\[40mm]} 
  \includegraphics[width=0.3\textwidth]{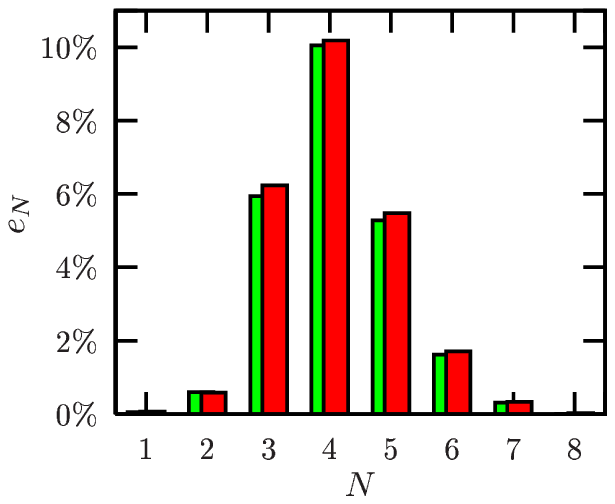}
  \quad\quad \parbox[b]{2mm}{(b)\\[40mm]} 
  \includegraphics[width=0.3\textwidth]{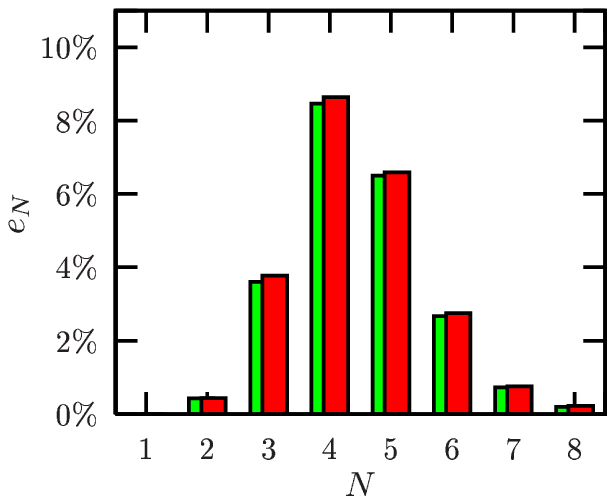}
  \quad\quad \parbox[b]{2mm}{(c)\\[40mm]} 
  \includegraphics[width=0.3\textwidth]{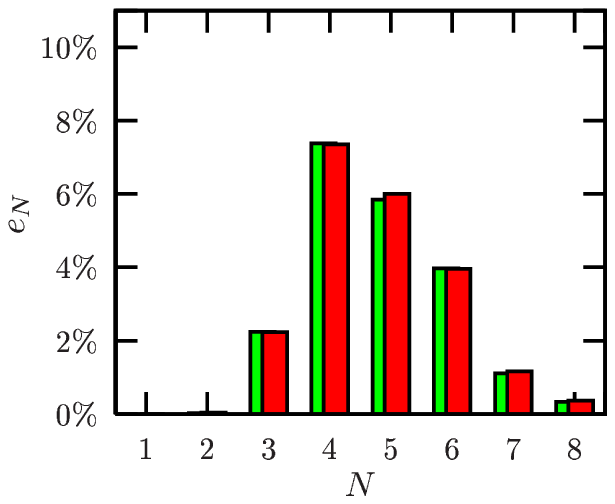}
  \]
  \caption[]{(color online) 
    Comparison of the statistical weight of coherent
    structures with $N$-fold symmetry in transiently chaotic time
    series of flows with different Reynolds numbers: (a)~$\Re=2200$,
    (b)~$\Re=2350$, (c)~$\Re=2500$. The red bars in the front are
    directly calculated from the simulation output, and the green bars
    in the background are the prediction of the Markov model. 
    \label{fig:probability}}
\end{figure*}

\section{Statistical analysis of the time series}
\label{sec:time-series}

Based on the time series $Z_N(t)$ we now explore the statistical
properties of the occurrence of coherent structures in pipe flow.
The aim of this statistical analysis is twofold: we want to see how
frequently structures of a certain multiplicity are present and we
want to study the extend to which a Markov approximation can describe
the switching between states.

\subsection{Probability distribution of coherent states}

\Fig{probability} shows the probabilities of detecting a coherent
state of $N$-fold symmetry in time series taken at different Reynolds
numbers \Re\ close to the transition to turbulence. For $\Re=2200$
about $24$\% of all cross sections fall into the categories $N=3$,
$4$, $5$, and $6$. For $\Re=2500$, the fraction decreases slightly to
about $20\%$. This high fraction explains the ease with which coherent
structures were picked out of experimental cross sections
\cite{Hof2004}, and underlines their significance as building blocks
of the turbulence in the transition region.

With increasing Reynolds number the weight of states with large $N$
increases. These structures are much closer to the walls where they
give rise to steeper gradients in radial and azimuthal direction and
consequently larger friction.  As these structures have more spatial
degrees of freedom, it is less likely that they appear in perfect
symmetry.  Hence, their correlators have smaller amplitudes, and it
would be interesting in forthcoming work to probe for the structures
with a localized correlator.

\begin{figure*}
  \[
  \parbox[b]{0mm}{(a)\\[38mm]} 
  \includegraphics[width=0.3\textwidth]{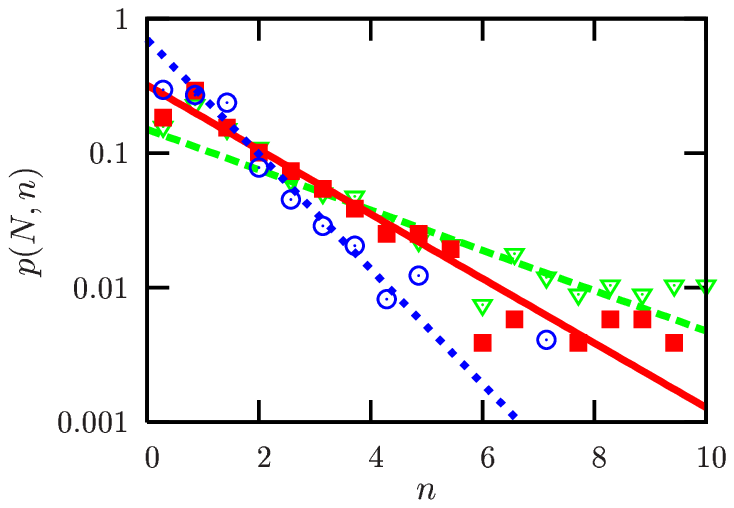}
  \quad\quad \parbox[b]{0mm}{(b)\\[38mm]} 
  \includegraphics[width=0.3\textwidth]{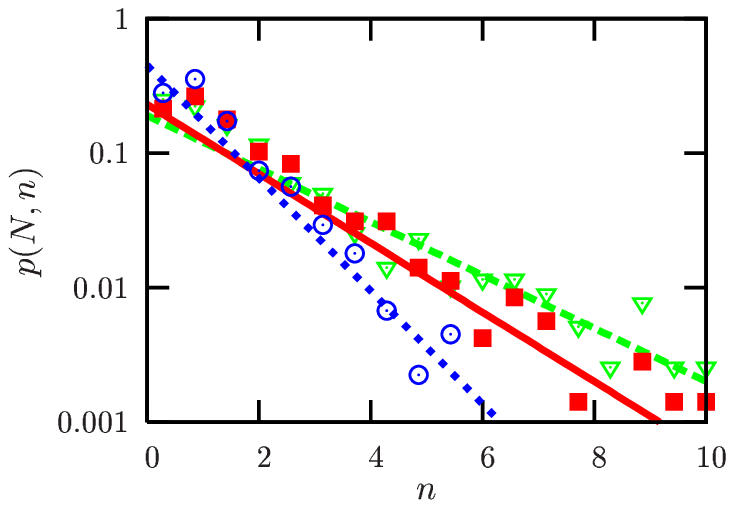}
  \quad\quad \parbox[b]{0mm}{(c)\\[38mm]} 
  \includegraphics[width=0.3\textwidth]{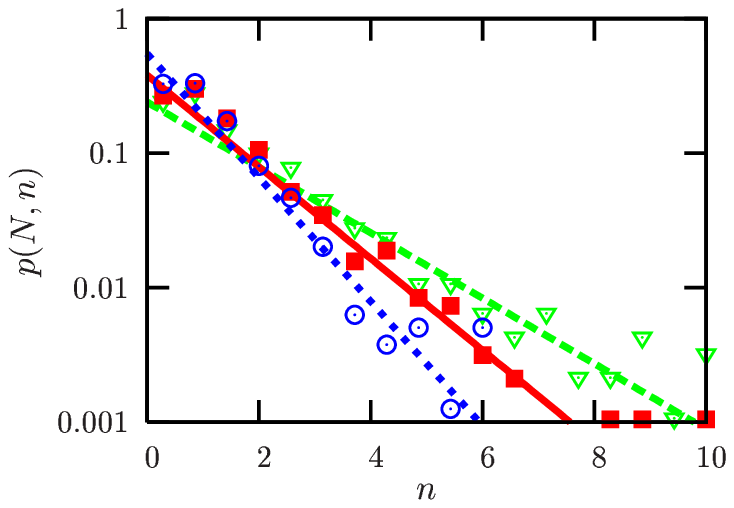}
  \]
  \caption[]{(color online) 
    Distribution of the persistence time of coherent
    structures with $N$-fold symmetry for the data also shown in
    \Fig{probability}.
    Following \Eq{persistence} the slope of the straight lines is 
    determined from the diagonal elements of the transfer matrix. 
    Time is measured in units of the 
    sampling time $\tau=1.4$. 
    ($N=4$: open blue circles and blue dotted line, 
     $N=5$: red filled boxes and solid red line, 
     $N=6$: open green triangles and green dashed line.) 
   \label{fig:lifetimedistri}
   }
\end{figure*}

\subsection{Markov model for transitions}

The typical persistence time of a pattern in \Fig{scalarmeasure} is
about $5$ to $10$ time units, and the transition between the
four-streak and six-streak state shown in \Fig{transition} takes about
one time unit.  When discretizing time in order to describe the
transitions between different patterns, the sampling time scale should
therefore not be much longer than about $5$.  Otherwise one misses
states.  On the other hand, if the time steps are much shorter than
unity, one begins to probe the continuity of the time evolution.  As
representative examples in this interval we explored the discrete
dynamics of discretized time sequences with a time spacing of
$\tau=1.4$ and of $\tau=2.4$.  Since different $\tau$ lead to results
which cannot be distinguished within our error margins, we will in the
following present data for $\tau=1.4$ only.

By considering the underlying flow at multiples of the time unit
$\tau$ its continuous dynamics is transformed into a discrete
time-series.
The conditional probability that one encounters an $N'$-streak state
in the following snapshot, when currently facing an $N$-streak state
defines a transition matrix $T_{N'\,N}$. Its indices $N$ and $N'$ take
the values $0$ (when there is no streak), and $N=2\ldots 8$ when $Z_N$
exceeds its threshold value.
For the Reynolds number $\Re=2200$ we find 
\begin{widetext}
\begin{equation}
  T_{}^{(\tau=1.4)}
  = \left(
  \begin{tabular}{l*{7}{@{\quad}l}}
    0.90 &0.26 &0.26 &0.27 &0.38 &0.55 &0.71 &1.00 \\
    0.00 &0.73 &0.00 &0.00 &0.00 &0.00 &0.00 &0.00 \\
    0.02 &0.00 &0.72 &0.01 &0.01 &0.01 &0.02 &0.00 \\
    0.04 &0.01 &0.01 &0.71 &0.03 &0.03 &0.07 &0.00 \\
    0.03 &0.00 &0.01 &0.01 &0.57 &0.03 &0.00 &0.00 \\
    0.01 &0.00 &0.00 &0.00 &0.01 &0.38 &0.04 &0.00 \\
    0.00 &0.00 &0.00 &0.00 &0.00 &0.00 &0.16 &0.00 \\
    0.00 &0.00 &0.00 &0.00 &0.00 &0.00 &0.00 &0.00 
%
%
  \end{tabular}
  \right)
\end{equation}
\end{widetext}
The columns of the matrices add up to $1$
because each state has to go to one of the eight admissible
states in the next time step,
\begin{equation}
   \sum_{N'=0,2,3\ldots 8} T_{N'\,N} = 1 \,.
\end{equation}
More than $10\,000$ independent snapshots ($17\,202$ for $\Re=2200$,
and more than $15\,000$ snapshots at the higher $\Re=2350$ and
$\Re=2500$) were analyzed in our statistics.  For the lowest Reynolds
number the $N=8$ class is observed only a single time, and it
immediately relaxed into the $N=0$ state (\cf rightmost column of
$T_{ij}$).  Despite its rare occurrence, the $N=8$ state is included
in the analysis because its statistical weight increases with Reynolds
number: It reaches $0.4 \%$ at $\Re=2500$.

\subsection{Invariant distribution and life time of coherent states}

To check that the Markovian dynamics generated by
the transition matrices faithfully represents the continuous dynamics,
we first calculate the invariant probability distribution $\vec{e}$, 
defined as the eigenvector to the eigenvalue 1, i.e.
\begin{equation}
\vec{e} = T \vec{e}\,.
\end{equation}%
\Fig{probability} shows that the $e_N$ faithfully reproduce the
relative frequencies in the original data.
A comparison of the histograms for the different \Re\ shows that the
number of visited coherent states and the complexity of the resulting
flow patterns increase when the Reynolds number increases.

Except for $N=8$ the highest transfer probabilities in each column
appear along the diagonal of $T$.  These elements describe the
persistence of flow patters from time step to time step.  Therefore, the
probability density function $p(N,n)$ to observe the pattern for
$n$ consecutive time steps scales like
\begin{equation}
   p(N,n) \sim (T_{NN})^n\,.
\label{eq:persistence}
\end{equation}
\Fig{lifetimedistri} shows data for the life time calculated from
  direct numerical simulation of the flow, together with the
  prediction from the Markov model, which is shown as straight lines
  in the semilogarithmic plot of lifetimes.
Since long persistence times are exponentially suppressed this
comparison requires very long time series to check the prediction
with reasonable statistical accuracy. Within these limitations there is
a very good agreement between the data and the prediction.

\begin{figure*}
\includegraphics[height=50mm]{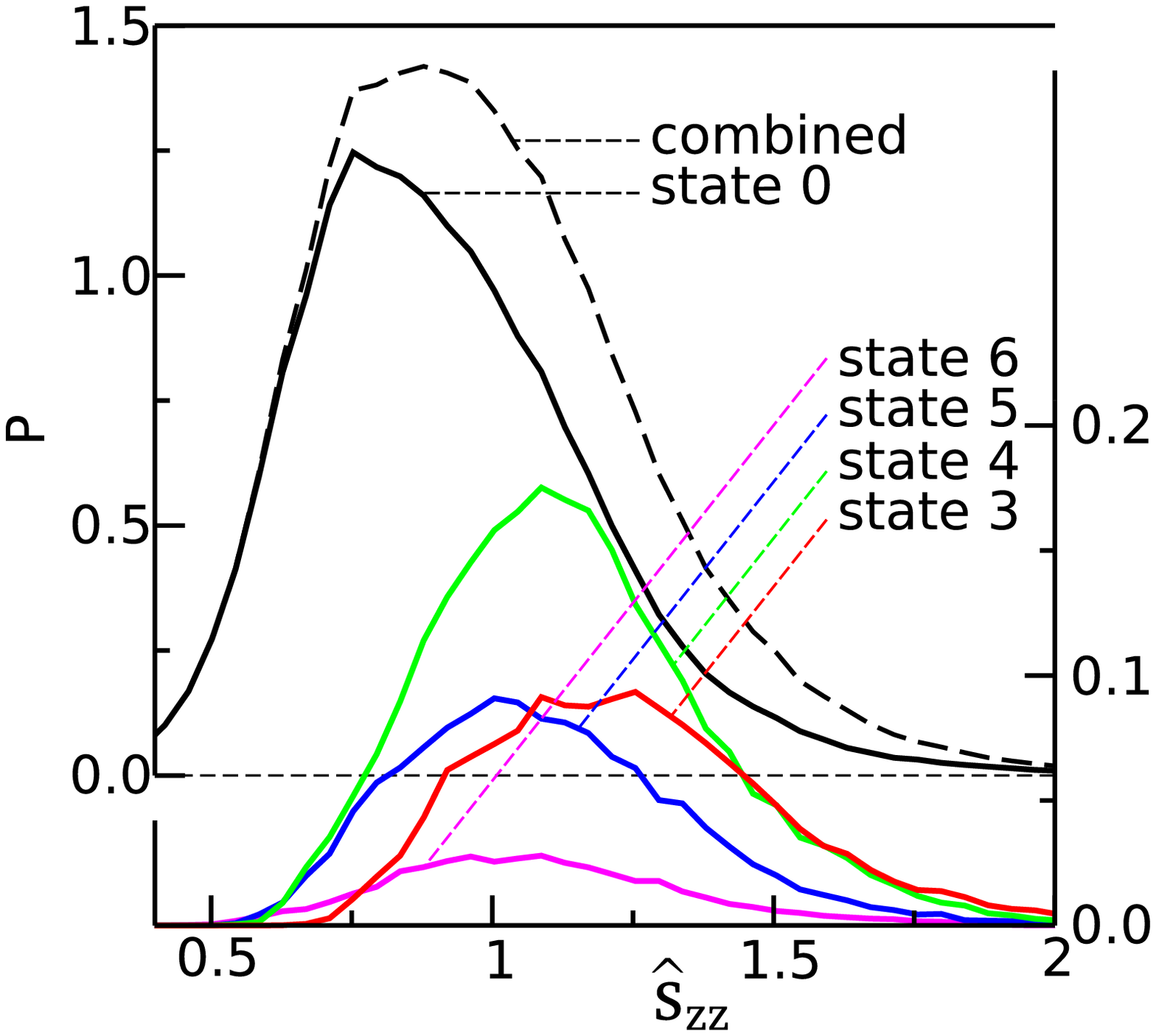}
\quad
\includegraphics[height=49mm]{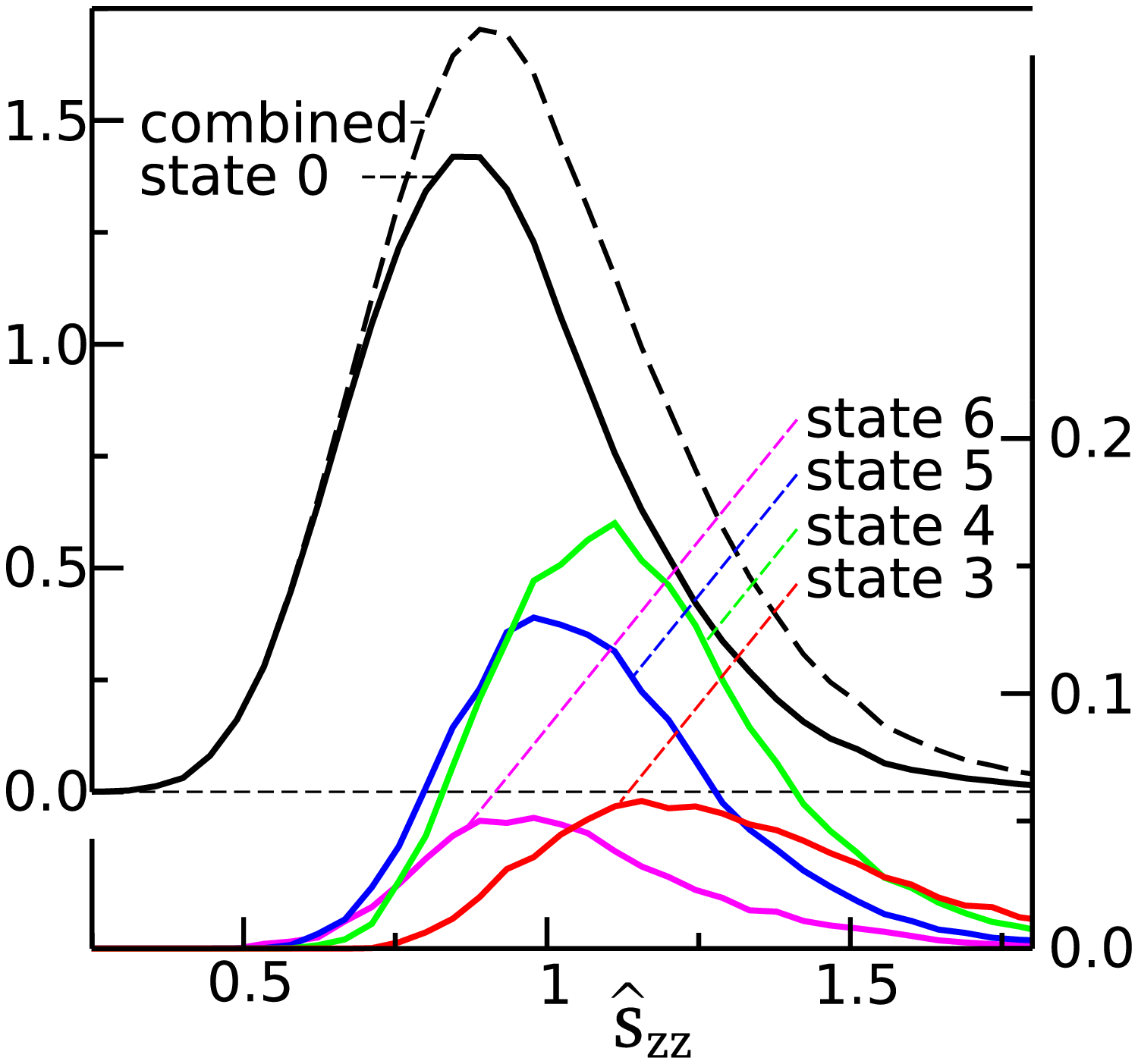}
\quad
\includegraphics[height=49mm]{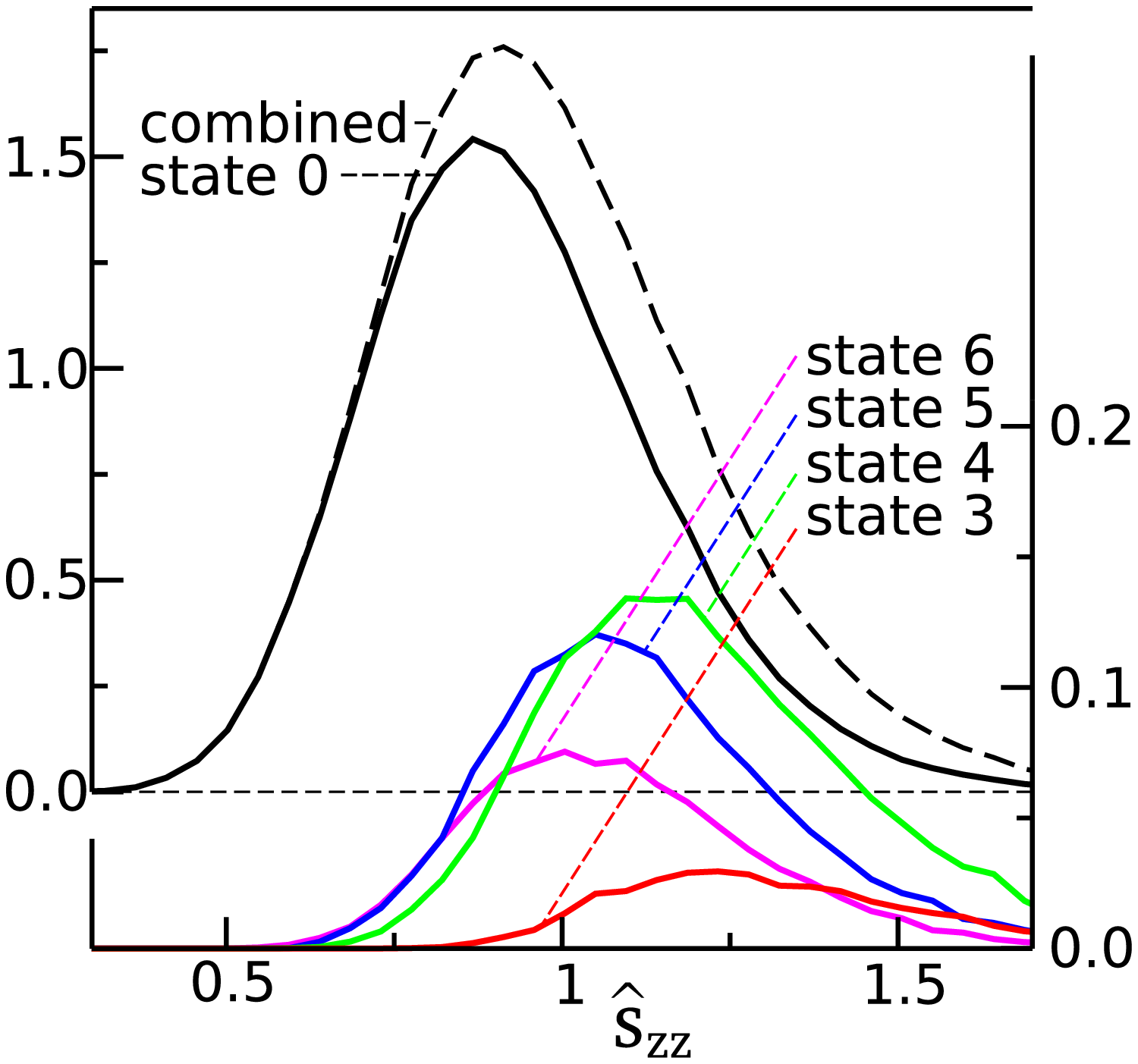}
\\[3mm]
\includegraphics[height=49mm]{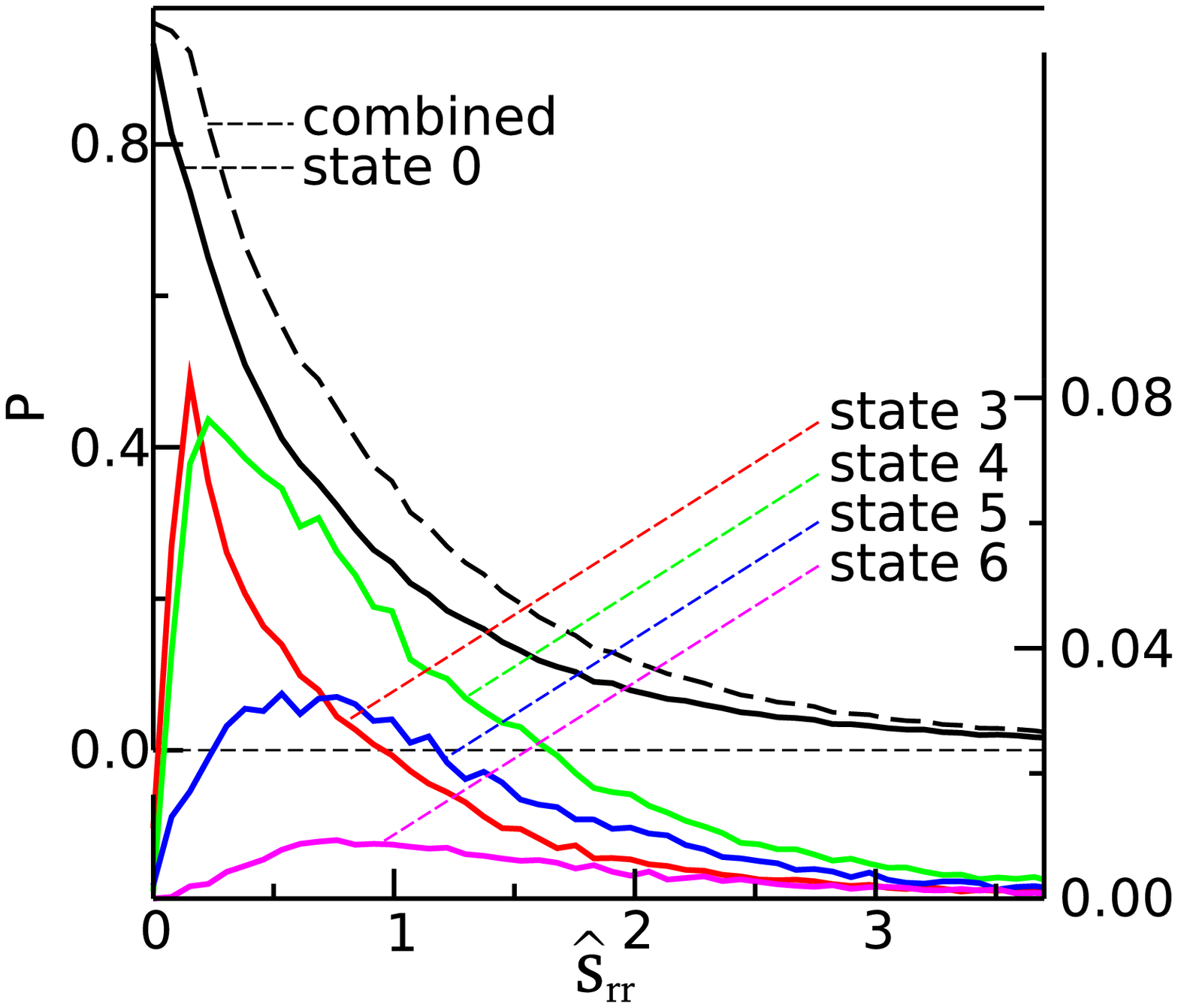}
\quad
\includegraphics[height=49mm]{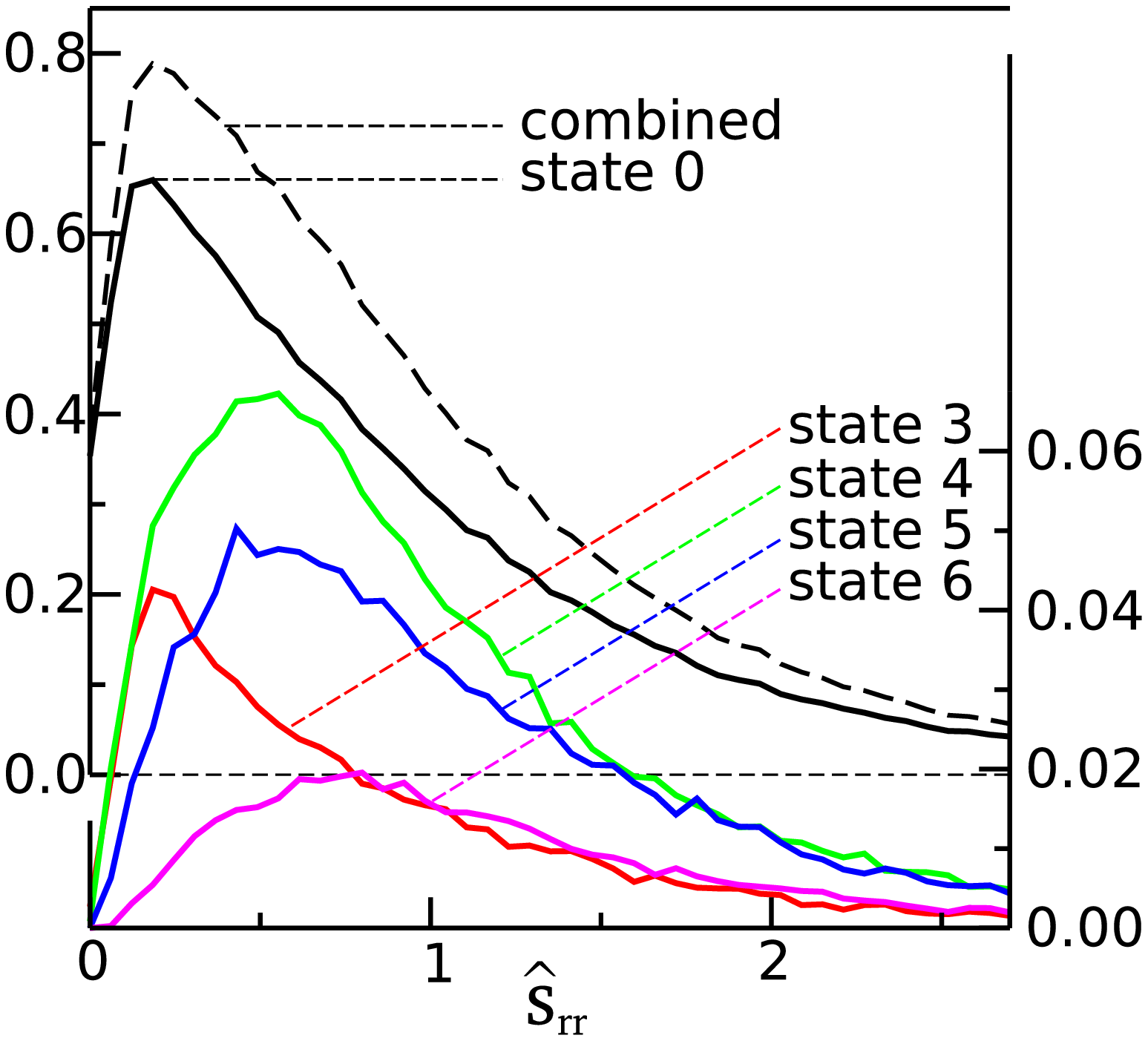}
\quad
\includegraphics[height=49mm]{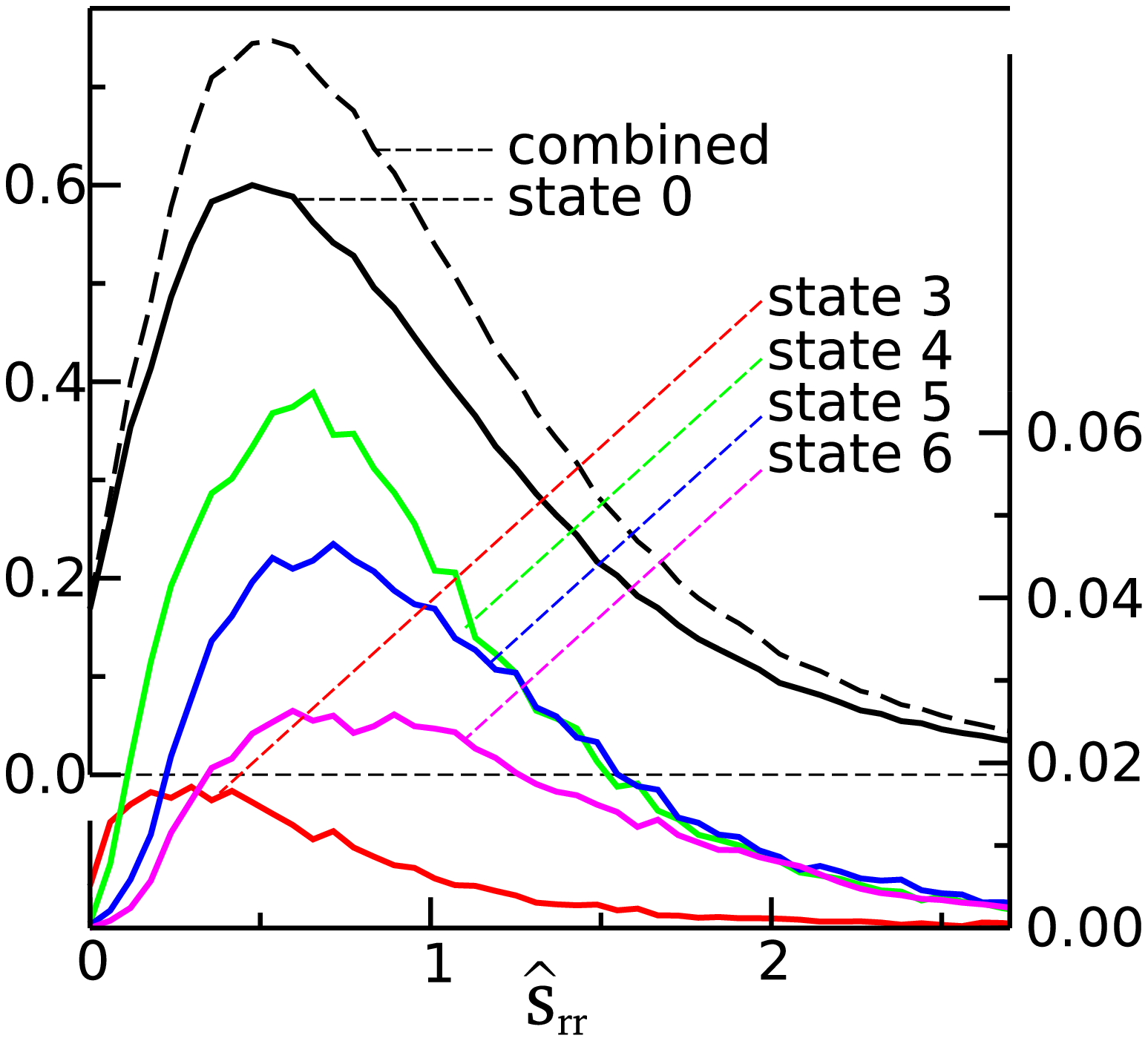}
\\[3mm]
\includegraphics[height=49mm]{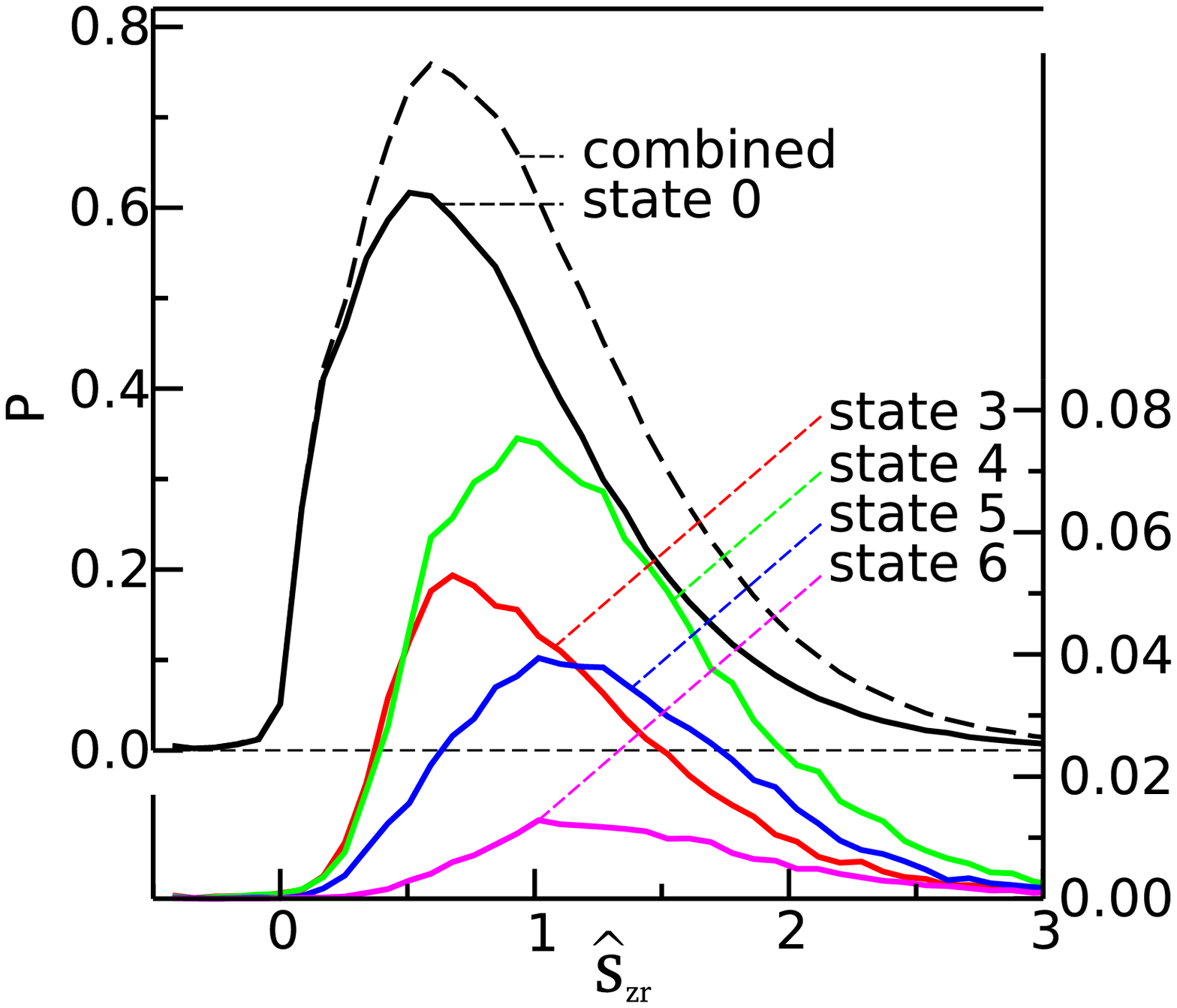}
\quad
\includegraphics[height=49mm]{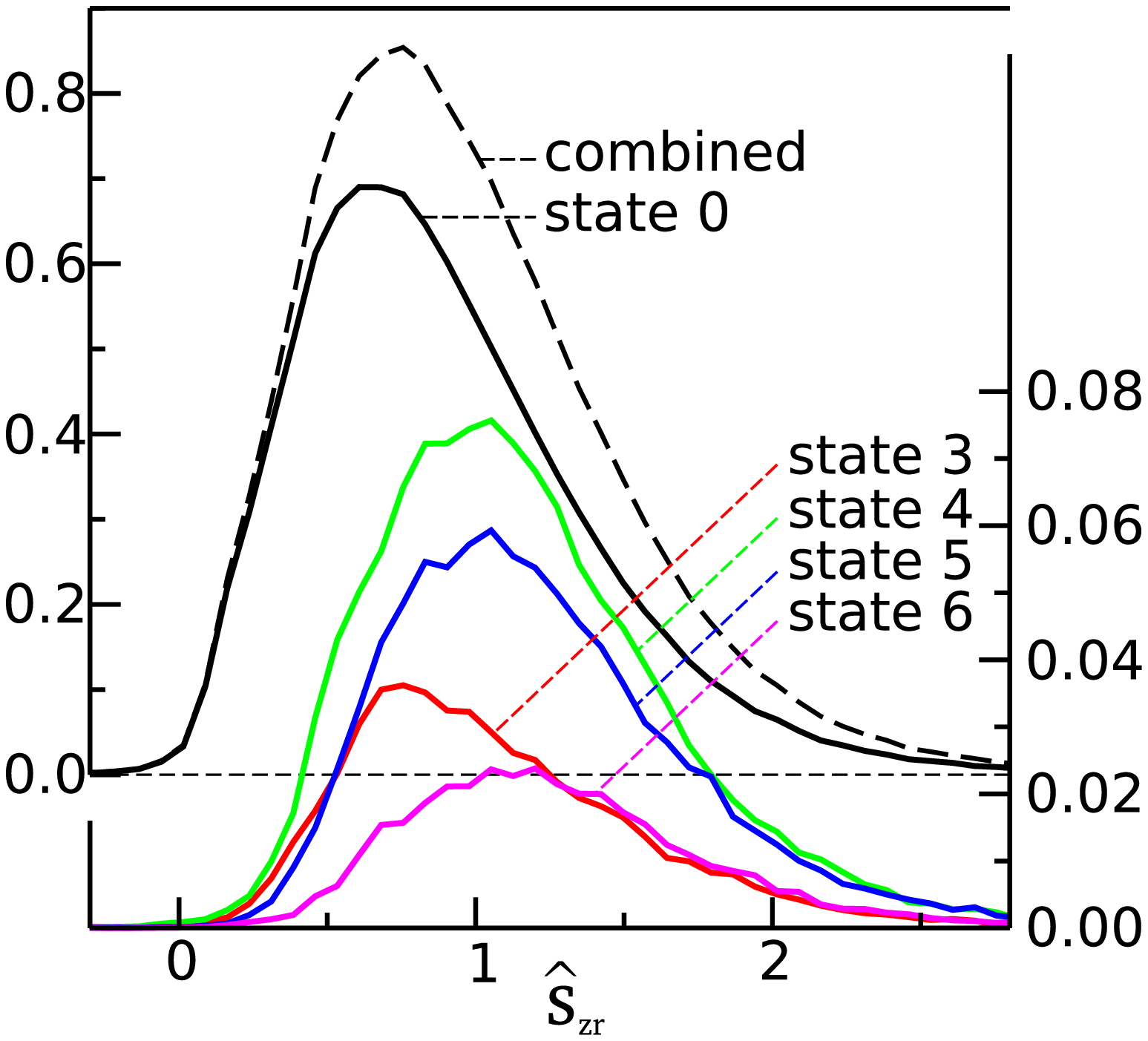}
\quad
\includegraphics[height=49mm]{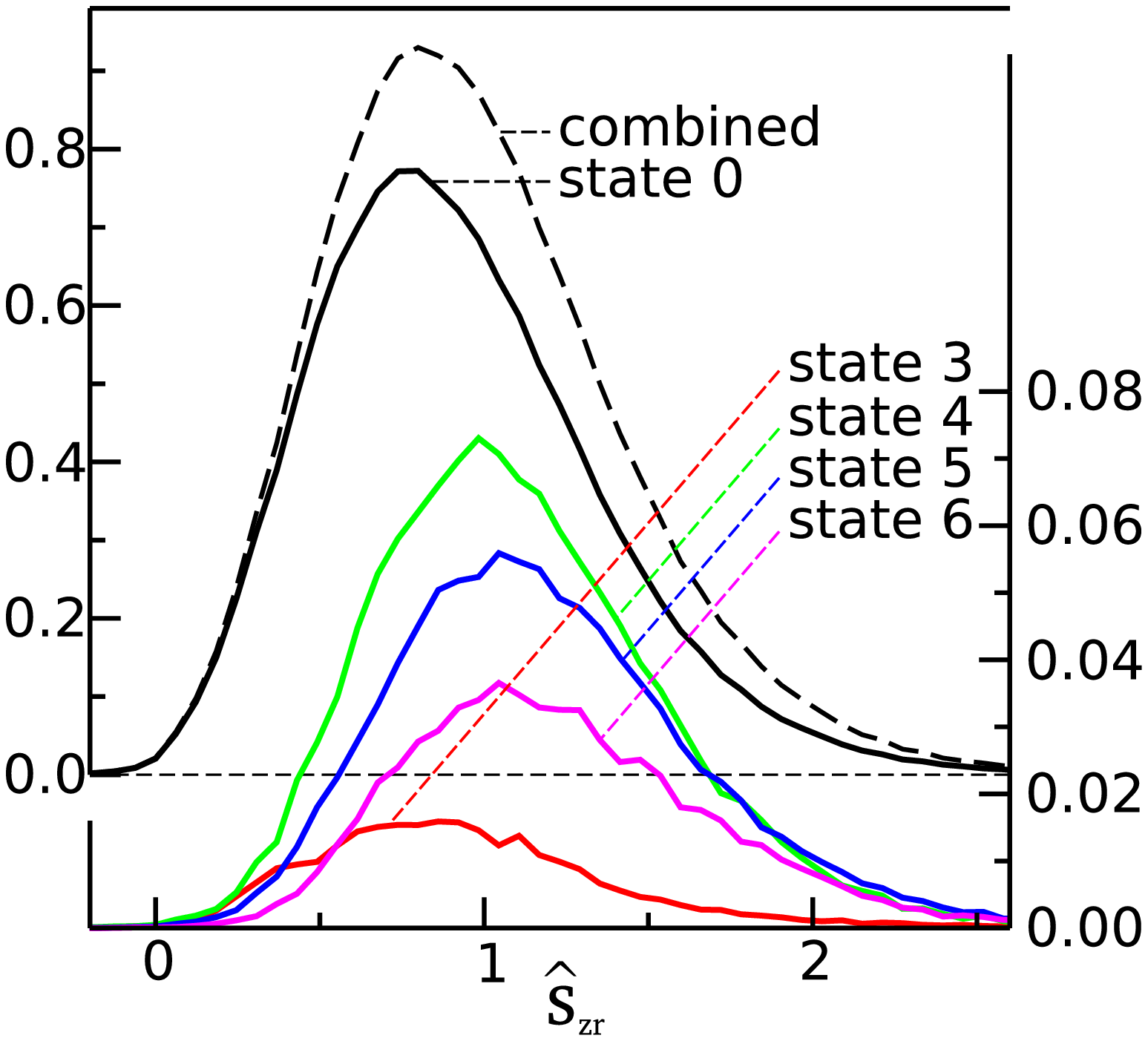}
\caption[]{(color online) 
  The axial velocity fluctuations (top), radial velocity
  fluctuations (middle), and radial momentum transport (bottom) for
  flows at Reynolds numbers $\Re=2200$ (left), $\Re=2350$ (center) and
  $\Re=2500$ (right).
  Different lines refer to the the overall
  time-averaged signal (dashed black, left axis), and the one averaged
  only over turbulent states, where no streaks are detected (solid
  black, left axes). The lines in the other colors give the respective
  contributions of states with a given number of streaks (right axis).
  The histograms are normalized with respect to their integral, \ie
  the overall distribution (dashed line) is normalized to unity, and
  all other distributions (given by solid lines) add up to the overall
  distribution. Their norm consequently amounts to the weights shown
  in the histograms in \Fig{probability}, and the position of their
  maxima indicate for which values they most strongly contribute to
  the overall signal.
\label{fig:fluctuations_axial}}
\end{figure*}

\section{Physical properties of detected states}
\label{sec:physProp}

The different flow patterns also affect the different velocity and
fluctuation statistics. As examples we consider the Reynolds stresses
$s_{zz}=\langle u_z u_z\rangle$, $s_{rr}=\langle u_r u_r\rangle$ and
$s_{zr}=\langle u_z u_r\rangle$.  Taking averages over $\phi$, but not
over time, provides probability distribution function (pdf) of
temporal variations of these quantities (dashed lines in
\Fig{fluctuations_axial} tagged as \textsf{combined}), as well as
conditional pdfs referring to states with a fixed number of streaks
(solid lines tagged as \textsf{state~3}\ldots\textsf{state~6}) and the
turbulent unstructured state (solid lines tagged as \textsf{state~0}).
The overall pdf can thus be decomposed into contributions of the
previously discussed high-symmetry coherent states and a turbulent
remainder (\textsf{state~0}).  To emphasize the role played by the
coherent states in changing the shape of the distribution of the
considered component of the Reynolds stress the abscissa is always
normalized to its overall temporal average. For instance, $s_{zz}$ is
normalized by its average $\overline{ s_{zz} }$, and the resulting
normalized stress is denoted $\hat{s}_{zz}=s_{zz}/\overline{ s_{zz}
}$. By definition the mean of the $\hat{s}_{zz}$ distribution is
therefore unity. However, the conditional pdfs for specific states
will in general have means different from one. If the mean is larger
than one, the state shows --- on average --- larger stress components
than the temporal average value of the component.  \Tab{meanValues}
lists both the absolute and the normalized mean values of all pdfs
shown in \Fig{fluctuations_axial}.

\begin{table}
\begin{tabular}{ll*{2}{l@{\;\;\;}l@{\quad\;}}l@{\;\;\;}l}
\hline\hline
& \Re & \multicolumn{2}{c}{$2200\qquad$\ } & \multicolumn{2}{c}{$2350\qquad$\ } & \multicolumn{2}{c}{$2500\qquad$\ }
\\
$ij$                    &      & $s_{ij}$   &$\hat{s}_{ij}$& $s_{ij}$ & $\hat{s}_{ij}$& $s_{ij}$ & $\hat{s}_{ij}$
\\[1mm]\hline
${zz}$                  & tot$\;$ & 
$1.05\;10^{-2}$ & $1.00$  & $1.03\;10^{-2}$  & $1.00$ & $0.96\;10^{-2}$   &  $1.00$
\\
                        & 0   & 
$0.98\;10^{-2}$ & $0.94$ & $0.98\;10^{-2}$  & $0.95$ & $0.92\;10^{-2}$    &  $0.95$
\\
                        & 3   & 
$1.34\;10^{-2}$ & $1.28$ & $1.36\;10^{-2}$  & $1.32$ & $1.30\;10^{-2}$    &  $1.35$
\\
                        & 4   & 
$1.22\;10^{-2}$ & $1.16$ & $1.21\;10^{-2}$  & $1.18$ & $1.19\;10^{-2}$    &  $1.24$
\\
                        & 5   & 
$1.17\;10^{-2}$ & $1.12$ & $1.14\;10^{-2}$  & $1.11$ & $1.11\;10^{-2}$    &  $1.15$
\\
                        & 6   & 
$1.16\;10^{-2}$ & $1.11$ & $1.10\;10^{-2}$  & $1.07$ & $1.06\;10^{-2}$    &  $1.10$
\\\hline
${rr}$                   & tot & 
$1.05\;10^{-4}$ & $1.00$ & $1.46\;10^{-4}$  & $1.00$ & $2.02\;10^{-4}$    &  $1.00$
\\
                        & 0   & 
$0.98\;10^{-4}$ & $0.94$ & $1.42\;10^{-4}$  & $0.97$ & $1.98\;10^{-4}$    &  $0.98$
\\
                        & 3   & 
$0.95\;10^{-4}$ & $0.90$ & $1.31\;10^{-4}$  & $0.90$ & $1.47\;10^{-4}$    &  $0.73$
\\
                        & 4   & 
$1.24\;10^{-4}$ & $1.18$ & $1.55\;10^{-4}$  & $1.06$ & $1.95\;10^{-4}$    &  $0.96$
\\
                        & 5   & 
$1.42\;10^{-4}$ & $1.35$ & $1.71\;10^{-4}$  & $1.17$ & $2.26\;10^{-4}$    &  $1.12$
\\
                        & 6   & 
$1.80\;10^{-4}$ & $1.71$ & $1.90\;10^{-4}$  & $1.29$ & $2.47\;10^{-4}$    &  $1.23$
\\\hline
${zr}$                   & tot & 
$4.73\;10^{-4}$ & $1.00$  & $5.62\;10^{-4}$  & $1.00$ & $6.51\;10^{-4}$    &  $1.00$
\\
                        & 0   & 
$4.29\;10^{-4}$ & $0.91$ & $5.26\;10^{-4}$  & $0.94$ & $6.19\;10^{-4}$    &  $0.95$
\\
                        & 3   & 
$5.34\;10^{-4}$ & $1.13$ & $6.23\;10^{-4}$  & $1.11$ & $6.32\;10^{-4}$    &  $0.97$
\\
                        & 4   & 
$6.12\;10^{-4}$ & $1.29$ & $6.74\;10^{-4}$  & $1.20$ & $7.41\;10^{-4}$    &  $1.14$
\\
                        & 5   & 
$6.51\;10^{-4}$ & $1.38$ & $7.09\;10^{-4}$  & $1.26$ & $7.95\;10^{-4}$    &  $1.22$
\\
                        & 6   & 
$7.23\;10^{-4}$ & $1.53$ & $7.41\;10^{-4}$  & $1.32$ & $8.24\;10^{-4}$    &  $1.27$
\\\hline\hline
\end{tabular}
\caption[mean values of pdfs]{
  The temporal mean (tot) of the Reynolds stresses
  $s_{zz}=\langle u_z u_z\rangle$, 
  $s_{rr}=\langle u_r u_r\rangle$, and
  $s_{zr}=\langle u_z u_r\rangle$ 
  in units of $4 \,\langle u_z \rangle^2$, and those of the
  corresponding conditional pdfs for disordered motion (\textsf{state~0}) and
  coherent states with $N=3\ldots 6$ streaks, respectively. 
  In addition also the corresponding relative values $\hat{s_{ij}}$
  are given  which are normalized with respect to the overall temporal
  mean of the considered component of the Reynolds stress.
  The related pdfs are shown in \Fig{fluctuations_axial}. 
  \label{tab:meanValues}}
\end{table}
\subsection{PDFs at fixed \Re}

From a physical point of view the interest of the decompositions of
the total pdf into conditional ones for turbulent and individual
coherent states lies in the insight it gives into how the coherent
states contribute to the exceptional statistics of fluctuations in
turbulent flow. We first consider the decomposition of the pdfs at
fixed \Re, \ie we discuss the trends in the mean of the data shown in
individual panels of \Fig{fluctuations_axial}.

On the average the detected coherent states generate much stronger
Reynolds stresses than those found for the unstructured turbulent
state $0$.  Consequently the coherent states shift the means and the
maxima of the combined pdf to slightly larger values.  Compared to the
pdf of \textsf{state~0} (dashed black line) the coherent states add a
fat tail to the combined pdf (black solid line) on the side of larger
values of the stresses.
In order to gain insight into the mutual importance of the different
states we discuss the trends in the mean and maxima as a function of
the number of streaks $N$.

The normalized stress $\hat{s}_{zz}$ characterizes the intensity of streak
structures in the flow field by estimating 
their downstream velocity.  The maxima and mean values of its pdf decrease
in the order $N=3$, $4$, $5$ and~$6$. 
This can be interpreted as follows:
The product of typical gradients of $u_z$ with the length scale over
which the gradients persist is of the order of magnitude of the
typical velocity fluctuation in downstream direction. Consequently,
the azimuthal components of the gradients of $u_z$ are of the same
order of magnitude in all coherent states, and their typical length
scale decreases like $N^{-1}$.

The radial component $\hat{s}_{rr}$ measures the typical fluctuations
of the radial velocity component, \ie it characterizes the strength of
the vortices. For this stress there also is a clear trend in the
position of the maxima and mean values with $N$, but with the sequence
reversed: the highest value for the maximum appears for $N=6$, and it
decreases towards $N=3$.  This finding suggests that stronger vortices
are needed to maintain the smaller streaks in coherent states with
larger $N$.

From a physical point of view the Reynolds stress $s_{zr}$ is the most
interesting of the three quantities.  After all, it reflects the
strength of the radial momentum transport. Hence it provides direct
insight in the friction factor in the turbulent flow
\cite{Eggels1994}, and it also immediately reflects the role of the
coherent states in the flattening of the laminar flow profile in
radial direction.
In view of the opposite scaling of the radial and axial velocity
components observed in $\hat{s}_{rr}$ and $\hat{s}_{zz}$,
respectively, its $N$ dependence results from a most subtle balance.
Indeed, the counteracting trends almost
cancel, leaving only a very weak decrease in the position of the maxima in
the sequence $N=6$, $5$, $4$ and $3$.

\begin{figure}
\[
\includegraphics[width=0.45\textwidth]{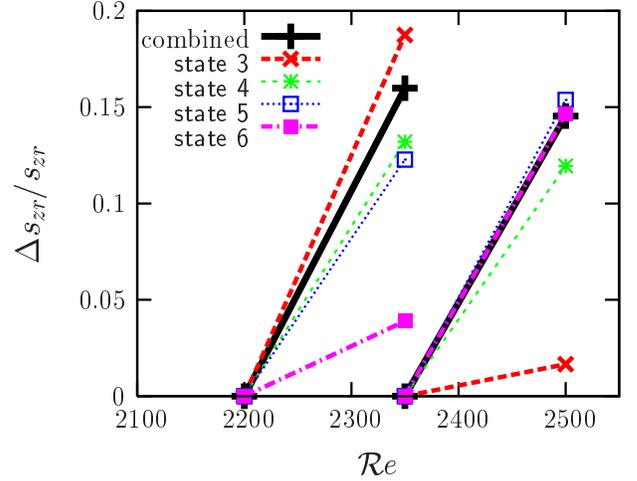}
\]
\caption[]{(color online) 
  The solid black line shows the contribution of the states with 
  steaks to the 
shift of the expectation value of ${s}_{zr}$ 
  when \Re\ is increased from
  $2200$ to $2350$ and from $2350$ to $2500$, respectively. 
  The shift results from those in the restricted pdfs 
  for states with $3$, $4$, $5$ and $6$ streaks, 
  which are shown by broken lines with colors matching the choice
  adopted in \Fig{fluctuations_axial}. 
  \label{fig:drifts}}
\end{figure}

\subsection{Drift of the mean with \Re}

In order to explore how the components of the Reynolds stress change
with Reynolds number, and which physical effects generate the observed
trends, one observes that the mean $\bar{x}$ of a combined pdf
$P(x) = \sum_N e_N P_N(x)$
with 
$\int\rmd x \, P(x) = 1$,
$\int\rmd x \, P_N(x) = 1$, and 
$\sum_N e_N = 1$
is the weighted average of the means $\bar{x}_N$ of the conditional
distributions $P_N(x)$,
\[
\bar{x} = \int\rmd x \, x P(x) 
        = \sum_N e_N \int\rmd x \, x P_N(x)
	= \sum_N e_N \bar{x}_N
\]
In \Fig{fluctuations_axial} the conditional pdfs $e_N P_N(x)$ are
plotted together with their sum $P(x)$ for $x=\hat{s}_{zz}$,
$\hat{s}_{rr}$, and $\hat{s}_{zr}$, respectively, and the abscissa is
scaled such that $\bar{x} = 1$.  The shift in the mean of $P(x)$
therefore arises as an average of the distance of the mean $\bar{x}_N$
from unity with weights $e_N$ previously discussed in the framework of
the Markov model (\cf\Fig{probability}).  There are two physical effects
underlying the observed changes in the statistics with \Re: 
(1) the change of the mean of conditional pdfs of the different states, and
(2) the change in the statistical weights of the states. 
We will disentangle these contributions now for the physically most
interesting case of $s_{zr}$.

Both visual inspection of the conditional pdf in
\Fig{fluctuations_axial} (bottom row), and the values of the
normalized mean values in \Tab{meanValues} show that there only is a
slight drift of the coherent states' pdfs with \Re. In contrast, as
observed upon discussing \Fig{probability} their weights show
pronounced changes.  The nontrivial evolution of their weights with
\Re\
suggests that the coherent states contribute to the change of the
overall mean mainly by the change of their statistical weights $e_N$.
This becomes particularly clear when plotting the relative change
$\Delta {s}_{zr} / {s}_{zr}$ of the position of the mean when
increasing \Re\ from $2200$ to $2350$ and from $2350$ to $2500$,
respectively (\Fig{drifts}). In the first interval this change is
dominated by the one of \textsf{state~3} while \textsf{state~6} hardly
contributes, and in the latter interval these two states take just the
opposite roles.

We thus conclude that our statistical analysis allows us to identify
the contributions of individual coherent states to the anomalous
statistics of turbulent pipe flow, and to disentangle the changes with
\Re\ into changes of the statistical weights of the states, and the
comparatively smaller ones due to the \Re-dependence of the properties
of individual states.  One can interpret these findings as a hint that
turbulent transients close to $\Re\gtrsim 2000$ are dominantly
influenced by coherent states with only few streaks. In contrast, at
higher \Re\ successively more coherent states with larger number of
streaks affect the time series.

\section{Discussion}
\label{sec:discussion}

In this section we want to summarize the results from the
present simulation of pipe flow and point to the parts
that could be useful in analyzing other shear flows as well.

The automatic detection algorithm for coherent states, which was first
used in \cite{Hof2004} and was expanded on here, is fairly robust.  It
can be generalized to other flows as well.  The algorithm
systematically searches for structures that show a symmetric azimuthal
arrangement of high-speed streaks along the wall which is
topologically very similar to the one observed in exact coherent
states reported in \cite{Faisst2003,Wedin2004,Hof2004}.
The detection is based on a Fourier-mode decomposition of the radial
velocity.  Because the detected states have the same symmetry
structure also in the other components of the velocity field, the
results should be robust against details of the implementation of the
detector.  Different projectors constructed along the line outlined in
\sect{detection} should lead to identical results.
For extensions to larger \Re\ it might, however, be valuable to
consider extensions to asymmetric expansions of the flow field, \eg by
using wavelet \cite{Daubechies1990,Gershenfeld2000} or Gabor
\cite{Groechenig2000,Feichtinger2006} representations to extract basic
units of coherent structures 
which contain only a 
single pair of vortices.

In principle, one can obtain more highly resolved and more accurate
information about the statistical properties of the flow by including
more degrees of freedom and subsequent extensions of the subspace of
projection.  In practice, however, the refinements are limited by the
available data set, because more degrees of freedom require many more
data points in order to guarantee statistically reliable results.

Irrespective of the chosen detector, the present method can be used to
\emph{quantitatively} analyze coherent structures in turbulent flow:
the automatic projectors give information about the probability to
observe certain coherent structures, their lifetimes, and the
transitions between these different states.
A number of observations can thus be made:

1. Despite being unstable the coherent states carry a considerable 
statistical weight of $\simeq 20$\%. 
This shows that even though the states are linearly unstable, they influence
the flow for a considerable part of its evolution. These numbers 
explain a posteriori why the states 
could be observed in the experiments by Hof et al \cite{Hof2004}. 

2. Upon increasing \Re\ from $2200$ to $2500$ the combined statistical
weight of all detected coherent states decreases only weakly.
However, there is a clear shift towards states with a larger number of
streaks.

3. Due to their prevalence the coherent states significantly influence
the turbulent dynamics at low \Re.  This opens a route to modelling
turbulence by exploring dynamical interconnections between coherent
states. To this end we considered the dynamics as a random walk
between a limited number of coherent states, and extracted the
transfer probabilities between states from the numerical time series.
The predictions of the Markov dynamics agree very well with the
numerically observed frequency of occurrence and the life time of the
coherent states.

4.  The decomposition of the Reynolds stresses into contributions
arising from irregular motion and contributions from coherent states
with three, four, five and six high-speeds streaks allowed us to study
the contribution of different structures to the radial momentum
transport.
Trends in the changes of the radial momentum transport with
\Re\ could be explained in terms of substantial changes of the
individual dynamical importance (statistical weights) of the states
while the properties of individual states change only slightly. Both
effects could be separated based on our statistical analysis.

We conclude that the methods presented in the present paper can be
used to \emph{quantitatively} analyze and describe turbulent dynamics
close to the transition to turbulence. Obviously, they can be extended
to projectors providing a still more detailed characterization of the
flow and used in other flows as well.  Since the approach does not
make use of specific features of our numerical setup, it should be
applicable to the analysis of numerical and experimental data alike.

\begin{acknowledgments}

  We are grateful to S.~Grossmann and A.~Jachens for helpful
  discussions and the \emph{Hessisches Hochleistungsrechenzentrum} in
  Darmstadt for computing time. This work was supported by the German
  Research Foundation.

\end{acknowledgments}


\end{document}